\documentclass[preprint2]{aastex}
\usepackage{amssymb,epsfig,wrapfig,graphics,aas_macros}
\usepackage[dvips]{color}
\usepackage[footnotesize]{caption}
\def\comment#1{{}}
\newlength{\cvindent}
\newlength{\cvhang}

\newlength{\refindent}
\def\lsim{\lower0.6ex\vbox{\hbox{$ \buildrel{\textstyle <}\over{\sim}\ $}}}
\def\gsim{\lower0.6ex\vbox{\hbox{$ \buildrel{\textstyle >}\over{\sim}\ $}}}

\begin{document}

\title{Section on Prospects for Dark Matter Detection of the White Paper on the Status and Future of Ground-Based TeV Gamma-Ray Astronomy}

\author{\normalsize
J.~Buckley (Wash.\ University in St. Louis, Physics Department and McDonnell Center for the Space Sciences),
E.A.~Baltz (KIPAC, Stanford University),
G.~Bertone (Institut d'Astrophysique de Paris, Université Pierre et Marie Curie),
K.~Byrum (Argonne National Laboratory),
B.~Dingus (Los Alamos National Laboratory),
S.~Fegan (University of California, Los Angeles),
F.~Ferrer (Wash.\ University in St. Louis, Physics Department and McDonnell Center for the Space Sciences),
P.~Gondolo (The University of Utah),
J.~Hall (Fermi National Accelerator Laboratory),
D.~Hooper (Fermi National Accelerator Laboratory),
D.~Horan (Argonne National Laboratory),
S.~Koushiappas (Brown University),
H.~Krawczynski (Wash.\ University in St. Louis, Physics Department and McDonnell Center for the Space Sciences),
S.~LeBohec (The University of Utah),
M.~Pohl (Iowa State University),
S.~Profumo (University of California, Santa Cruz),
J.~Silk (Oxford University),
T.~Tait (Argonne National Laboratory and Northwestern University),
V.~Vassiliev (University of California, Los Angeles)
R.~Wagner (Argonne National Laboratory),
S.~Wakely (The University of Chicago),
M.~Wood (University of California, Los Angeles),
and G.~Zaharijas (Argonne National Laboratory)}

\begin{abstract}

This is a report on the findings of the dark matter science working group for
the white paper on the status and future of TeV gamma-ray astronomy. The white
paper was commissioned by the American Physical Society, and the full white
paper can be found on astro-ph (arXiv:0810.0444). This detailed section
discusses the prospects for dark matter detection with future gamma-ray
experiments, and the complementarity of gamma-ray measurements with other
indirect, direct or accelerator-based searches.
We conclude that any comprehensive search for dark matter 
should include gamma-ray observations, both to 
identify the dark matter particle (through the characteristics of the
gamma-ray spectrum) and to measure the distribution of dark matter in
galactic halos.
\vspace*{1.5cm}
\end{abstract}

\maketitle

\section{Introduction}

In the last decade, a standard cosmological picture of the universe (the
$\Lambda$CDM cosmology) has emerged, including a detailed breakdown of the main
constituents of the energy-density of the universe.  This theoretical framework
is now on a firm empirical footing, given the remarkable agreement of a diverse
set of astrophysical data \cite{Spergel:2006hy,Percival:2006gt}.  In the
$\Lambda$CDM paradigm, the universe is spatially flat and its energy budget is
balanced with $\sim$4\% baryonic matter, $\sim$26\% cold dark matter (CDM) and
roughly 70\% dark energy.

\begin{figure*}[tb]
\label{fig:halos}
\begin{center}
\includegraphics[width=0.96\textwidth]{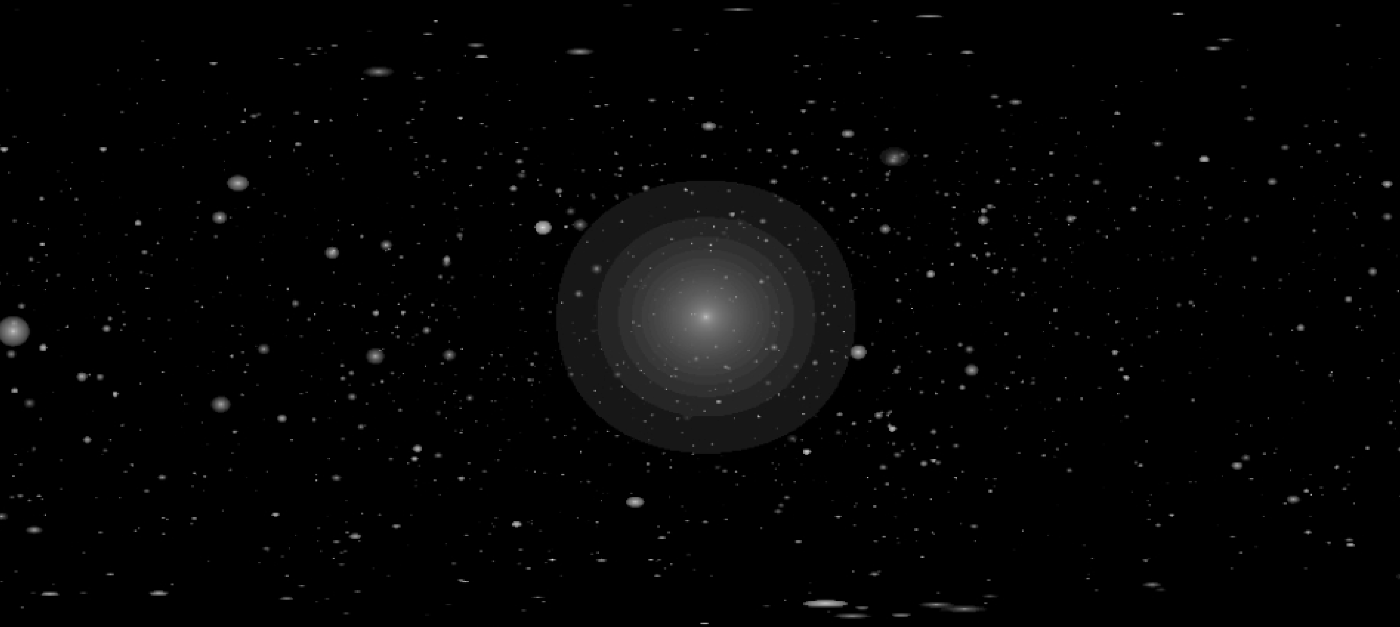}
\end{center}
\caption{Simulated appearance of the gamma-ray sky from neutralino annihilation
in the galactic halo plotted as the intensity in galactic coordinates
\cite{baltz_p5_06}.  The galactic center appears as the bright object at the
center of the field of view.  If the sensitivity of a future ACT experiment
were high enough, a number of the other galactic substructures visible in this
figure could be detected with a ground-based gamma-ray experiment.}

\end{figure*}

While the dark matter has not been directly detected in laboratory experiments,
the gravitational effects of dark matter have been observed in the Universe on
all spatial scales, ranging from the inner kiloparsecs of galaxies out to the
Hubble radius.  The Dark Matter (DM) paradigm was first introduced by Zwicky
~\cite{Zwicky1933} in the 1930s to explain the anomalous velocity dispersion in
galaxy clusters. 

In 1973, Cowsik and McClelland \cite{CowsikMcClelland(1973)} proposed that
weakly-interacting massive neutrinos could provide the missing dark matter
needed to explain the virial mass discrepancy in the Coma cluster. However,
since neutrinos would be relativistic at the time of decoupling, they would
have a large free-streaming length. While neutrino dark matter would provide an
explanation for structure on the scale of clusters, this idea could not explain
the early formation of compact halos that appear to have seeded the growth of
smaller structures, such as galaxies.

This observation motivated the concept of cold dark matter (CDM) consisting of
weakly interacting massive particles (WIMPs) with rest energy on the order of
100 GeV that were nonrelativistic (cold) at the time of decoupling.  CDM would
first form very small, dense structures that coalesced into progressively
larger objects (galactic substructure, galaxies, then galaxy clusters and
superclusters) in a bottom-up scenario known as hierarchical structure
formation.  A plethora of diverse observations suggests the presence of this
mysterious matter: gravitational lensing, the rotation curves of galaxies,
measurements of the cosmic microwave background (CMB), and maps of the
large-scale structure of galaxies.

Measurements of the CMB have been the key to pinning down the cosmological
parameters; the angular distribution of temperature variations in the CMB
depends on the power spectrum of fluctuations produced in the inflationary
epoch and subsequent acoustic oscillations that resulted from the interplay of
gravitational collapse and radiation pressure.  These acoustic peaks contain
information about the curvature and expansion history of the universe, as well
as the relative contributions of baryonic matter, dark matter and dark energy.
Combined with measurements of the large-scale distribution of galaxies, as
mapped by the Sloan Digital Sky Survey (SDSS) and the 2dF Galaxy Redshift
survey, these data can be well described by models based on single field
inflation.

Observations of galactic clusters continue to be of central importance in
understanding the dark matter problem.  Recent compelling evidence for the
existence of particle dark matter comes from the analysis of a unique cluster
merger event 1E0657-558 \cite{clowe06}.  Chandra observations reveal that the
distribution of the X-ray emitting plasma, the dominant component of the
visible baryonic matter, appears to be spatially segregated from the
gravitational mass (revealed by weak lensing data).  This result provides
strong evidence in favor of a weakly-interacting-particle dark matter, while
contradicting other explanations, such as modified gravity.

The primordial abundances of different particle species in the Universe are
determined by assuming that dark matter particles and all other particle
species are in thermal equilibrium until the expansion rate of the Universe
dilutes their individual reaction rates.  Under this assumption (which provides
stunningly accurate estimates of the abundance of light elements and
standard-model particles), particles that interact weakly fall out of
equilibrium sooner, escaping Boltzmann suppression as the temperature drops,
and hence have larger relic abundances in the current universe.  While a
weakly-interacting thermal relic provides an appealing and well-constrained
candidate for the dark matter, nonthermal relics such as axions or gravitinos,
resulting from the decay of other relics, can also provide contributions to the
total matter density or even provide the dominant component of the dark matter.
Just as there is an unseen component of the universe required by astrophysical
observations, there are compelling theoretical arguments for the existence of
new particle degrees of freedom in the TeV to Planck scale energy range.  In
particle physics, a solution to the so-called hierarchy problem (the question
of why the expected mass of the Higgs particle is so low) requires new physics.
An example is provided by supersymmetry, a symmetry in nature between Fermions
and bosons, where the supersymmetric partners of standard model particles lead
to cancellations in the radiative corrections to the Higgs mass.  The hierarchy
problem in particle physics motivates the existence of new particle degrees of
freedom in the mass range of ~100~GeV to TeV scale.  It is a remarkable
coincidence that if dark matter is composed of a weakly interacting elementary
particle with an approximate mass of this order (i.e., on the scale of the weak
gauge bosons $\sim 100$ GeV), one could naturally produce the required
cosmological density through thermal decoupling of the DM component.  To make a
viable candidate for the dark matter, one more ingredient is required; the
decay of such a particle must be forbidden by some conserved quantity
associated with an, as yet, undiscovered symmetry of Nature so that the
lifetime of the particle is longer than the Hubble time.

In supersymmetry, if one postulates a conserved quantity arising from some new
symmetry (R-parity), the lightest supersymmetric particle (LSP) is stable and
would provide a natural candidate for the dark matter.  In fact, R-parity
conservation is introduced into supersymmetry not to solve the dark matter
problem, but rather to ensure the stability of the proton.  In many regions of
supersymmetric parameter space, the LSP is the neutralino, a Majorana particle
(its own antiparticle) that is the lightest super-symmetric partner to the
electroweak and Higgs bosons.  

For a subset of the supersymmetric parameter space, these particles could be
within the reach of experimental testing at the Large Hadron Collider (LHC) (if
the rest mass is below about 500~GeV) \cite{baltz06}  or current or future
direct detection experiments XENON-I,II \cite{aprile05}, GENIUS
\cite{klapdor02,klapdor05} ZEPLIN-II,III,IV \cite{bisset07},
SuperCDMS\cite{akerib05}, and EDELWEISS-I,II\cite{sanglard07} (if the nuclear
recoil cross-section is sufficiently large).  While it is possible that the LHC
will provide evidence for supersymmetry, or that future direct detection
experiments will detect a clear signature of nuclear-recoil events produced by
dark matter in the local halo, {\emph{gamma-ray observations provide the only
avenue for measuring the dark matter halo profiles and illuminating the role of
dark matter in structure formation.}} 

Neutralinos could also be observed through other indirect astrophysical
experiments searching for by-products of the annihilation of the lightest
supersymmetric particle, such as positrons, low-energy antiprotons, and
high-energy neutrinos.  Since positrons and antiprotons are charged particles,
their propagation in the galaxy suffers scattering off of the irregular
inter-stellar magnetic field and hides their origin.  Electrons with energy
above $\sim$10~GeV suffer severe energy losses due to synchrotron and
inverse-Compton radiation, limiting their range to much less than the distance
between Earth and the galactic center.  However, cosmic-ray observations could
provide evidence for local galactic substructure through characteristic
distortions in the energy spectra of these particles.  Detection of electrons
from dark matter annihilation thus depend critically on large uncertainties in
the clumpiness of the local halo.  Neutrinos would not suffer these
difficulties and, like photons, would point back to their sources.  But given
the very low detection cross section compared with gamma-rays, the effective
area for a $\sim$km$^3$ neutrino experiment is many orders of magnitude smaller
than for a typical ground-based gamma-ray experiment.  While detection of
neutrinos directly from discrete sources (e.g., the Galactic center) would be
difficult for the current generation of neutrino detectors there is a
reasonable prospect for detection of neutrinos from WIMPs s in the local halo
that are captured by interactions with the earth or sun where they might have
sufficient density to give an observable neutrino signal.  {Compared with all
other detection techniques (direct and indirect), $\gamma$-ray measurements of
dark-matter are unique in going beyond a detection of the local halo to
providing a measurement of the actual distribution of dark matter on the sky.
Such measurements are needed  to understand the nature of the dominant
gravitational component of our own Galaxy, and the role of dark matter in the
formation of structure in the Universe.}

In other regions of supersymmetric parameter space, the dark matter particle
could be in the form of a heavy scalar like the sneutrino, or Rarita-Schwinger
particles like the gravitino.  In general, for gravitino models, R-parity need
not be conserved and gravitinos could decay very slowly (with a lifetime on the
order of the age of the universe) but could still be visible in gamma-rays
\cite{ibarratran08}.  Supersymmetry is not the only extension to the standard
model of particle physics that provides a dark matter candidate, and there is
no guarantee that even if supersymmetry is discovered it will provide a new
particle that solves the dark matter problem.  Other extensions of the standard
model involving TeV-scale extra dimensions, include new particles in the form
of Kaluza-Klein partners of ordinary standard-model particles.  The lightest
Kaluza-Klein particle (LKP) could be stable and hence provide a candidate for
the dark matter if one invokes an absolute symmetry (KK parity conservation)
resulting from momentum conservation along the extra dimension.  The mass of
the lightest Kaluza-Klein particle (e.g, the $B^{(1)}$ particle corresponding
to the first excitation of the weak hypercharge boson) is related to the
physical length scale of the extra dimension and could be on the TeV-scale (but
not much smaller) and provide a viable CDM candidate.  The $B^{(1)}$ is
expected to annihilate mainly to quarks or charged leptons accompanied by an
internal bremsstrahlung photon by the process $B^{(1)}+B^{(1)}\rightarrow l^+ +
l^- + \gamma$ \cite{bergstrom06}.  The high energy of the LKP ($\gsim$1 TeV),
and very-hard spectrum gamma-ray production make ground-based gamma-ray and
high-energy cosmic-ray electron measurements promising avenues for discovery. 

As an interesting aside, TeV-scale extra dimensions may also manifest
themselves in a dispersion in the propagation velocity of light in
extragalactic space \cite{amelino98}.  Observations of the shortest flares, at
the highest energies from the most distant objects can place tight constraints
on theories with large extra dimensions.  Such constraints have already been
produced by TeV measurements \cite{biller99} and could be dramatically improved
with a future higher-sensitivity gamma-ray instrument, capable of detecting
shorter flares from distant AGNs and GRBs.  {Thus, ground-based TeV gamma-ray
astronomy probes TeV-scale particle physics both by providing a possible avenue
for detection of a Kaluza-Klein particle and by constraining the the
TeV$^{-1}$-scale structure of space-time from gamma-ray propagation effects.}

A new class of theories (the so-called ``little Higgs'' or LH models) has been
proposed to extend the standard model to the TeV scale and offer an explanation
for the lightness of the Higgs.  The LH models predict a light (possibly
composite) Higgs boson as well as other TeV-scale particles that could provide
candidates for the dark matter in the $\sim$100~GeV or $\gsim$500~GeV mass
range \cite{birkedal06}.  However, only a small subset of the LH models have
weak-scale masses and interactions together with a symmetry principle that
protects the stability of the particle on a lifetime comparable to the age of
the universe.  In fact, for the composite Higgs, the particles (like their
analog, the neutral pion) could decay with relatively short lifetimes.  Still,
this class of models (like other new physics at the TeV scale) could provide a
viable dark matter candidate with an observable gamma-ray signature.

\begin{figure}[tb]
\label{fig:bringmann}
\includegraphics[width=0.48\textwidth]{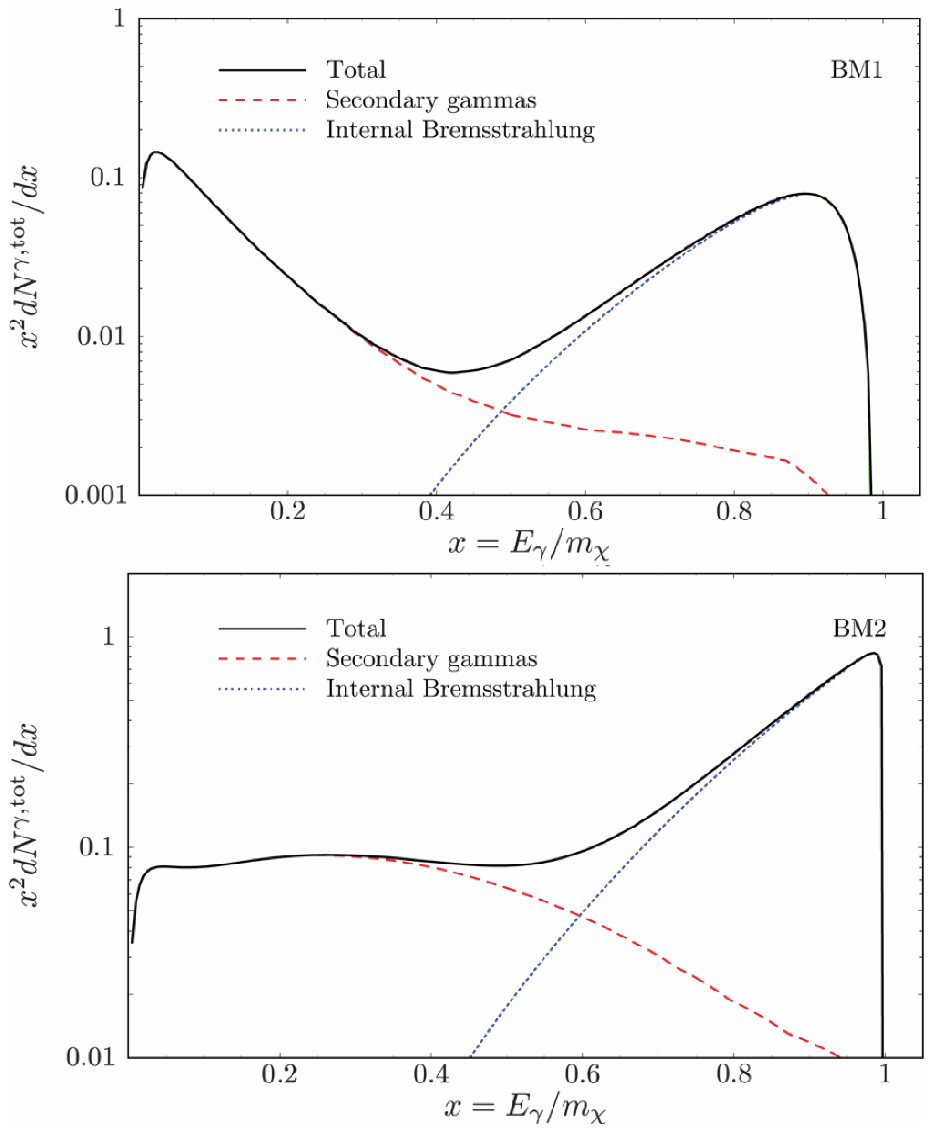}
\caption{Continuum emission from neutralino annihilation from mSUGRA models.}
\end{figure}

\begin{figure}[tb]
\includegraphics[width=0.48\textwidth]{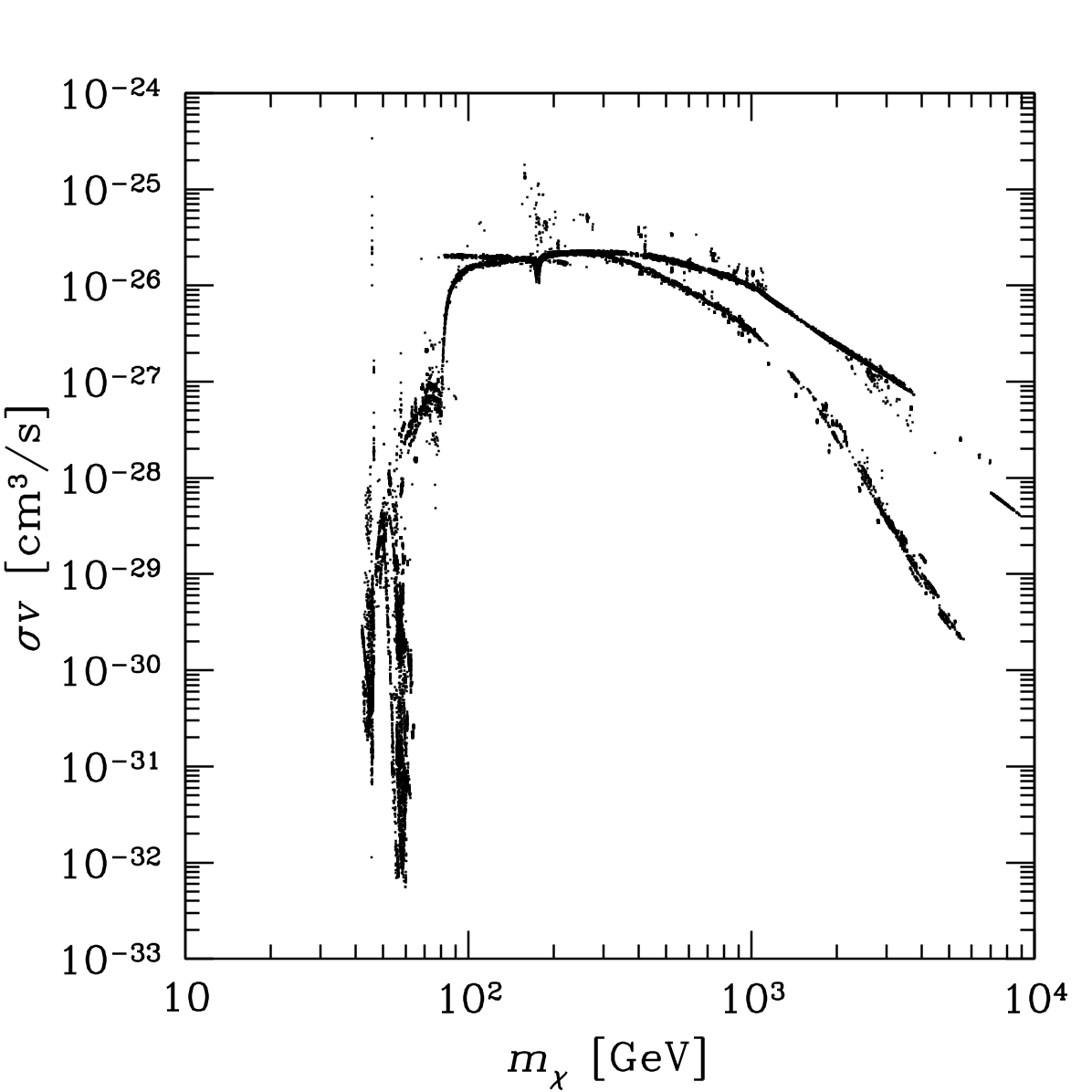}
\caption{
Scatter plot of neutralino annihilation cross section versus neutralino
mass for supersymmetric models that satisfy accelerator and WMAP constraints.
A typical cross-section (assumed in our estimates) is $\sigma v \approx
2\times 10^{-26}{\rm cm}^3 {\rm s}^{-1}$.
\label{fig:gondolosigma}}
\end{figure}

The recent discoveries of neutrino mass from measurements of atmospheric and
solar neutrinos may also have a bearing on the prospects for gamma-ray
detection of dark-matter.  While the primordial density of light standard-model
(SM) neutrinos $\nu_e$, $\nu_\mu$ and $\nu_\tau$ will provide a very small
hot-dark-matter contribution to the energy budget of the universe, they are
ruled out as candidates for the CDM component needed to explain structure
formation.  However, a new heavy neutrino (or the superpartner thereof) may
provide a viable candidate for the CDM.  Krauss, Nasri and Trodden
\cite{kraussnasri} proposed that a right-handed neutrino with TeV mass could
play a role in giving masses to otherwise massless standard model neutrinos
through high-order loop corrections.  This model is a version of the Zee model
\cite{zee80} that has been successfully applied to results on solar and
atmospheric neutrino observations to explain the observed parameters of the
mass and mixing matrix.  A discrete $Z_2$ symmetry, and the fact that the
right-handed Majorana neutrino $N_R$ is typically lighter than the charged
scalars in the theory, make the massive neutrino stable, and a natural dark
matter candidate \cite{baltzbergstrom}.  Direct annihilation to a gamma-ray
line $N_R N_R\rightarrow \gamma\gamma$ with a cross-section $\langle\sigma_{N_R
N_R\rightarrow \gamma\gamma} v\rangle \approx 10^{-29}{\rm cm}^{3}{\rm s}^{-1}$
is at the limit of detectability and direct annihilation to charged leptons is
also expected to give a very small cross-section.  However,
\cite{baltzbergstrom} have shown that internal bremsstrahlung can give rise to
an observable gamma-ray continuum from decays to two leptons and a gamma-ray
$N_R N_R\rightarrow l^+ l^- \gamma$.  The three-body final state gives rise to
a very hard spectrum that peaks near the $N_R$ mass, then drops precipitously.
Unlike direct annihilation to leptons, this non-helicity-suppressed process can
have a large cross-section, with an annihilation rate a factor of $\alpha/\pi$
(where $\alpha$ is the fine structure constant) times the annihilation rate at
freeze-out (with cross section $\langle \sigma v\rangle \approx 3\times
10^{-26}{\rm cm}^3{\rm s}^{-1}$), and orders of magnitude lager than the
helicity-suppressed two-body $N_R N_R\rightarrow l^+ + l^-$ rate typically
considered in the past \cite{baltzbergstrom}.

Recently, Bringmann, Bergstr\"om and Edsj\"o \cite{bringmann07} have pointed
out that internal-bremsstrahlung process could also play a role in neutralino
annihilation, and in some cases result in a large enhancement in the continuum
gamma-ray signal for certain model parameters.  Fig.~\ref{fig:bringmann} shows
the continuum emission from neutralino annihilation from mSUGRA models with
particularly pronounced IB features, that could be observed in the gamma-ray
spectrum.  {There are a number of different particle physics and astrophysical
scenarios that can lead to the production of an observable gamma-ray signal
with a spectral form that contains distinct features that can be connected,
with high accuracy, to the underlying particle physics}.

In what follows, we focus on predictions for the neutralino.  While we show
detailed results for the specific case of SUSY models and the neutralino, for
any theory with a new weakly interacting thermal relic (e.g., the LKP) the
model parameter space is tightly constrained by the observed relic abundance
and hence the results for the overall gamma-ray signal level are fairly generic
for any WIMP candidate.  In the case of neutralino dark matter, the
cross-sections for annihilation have been studied in detail by a number of
groups.  Fig.~\ref{fig:gondolosigma} shows the cross-section calculated for a
range of parameters in supersymmetric parameter space as a function of mass.
Only points that satisfy accelerator constraints and are compatible with a
relic abundance matching the WMAP CMB measurements are shown.  At high
energies, the neutralino is either almost purely a Higgsino (for mSUGRA) or
Wino (for anomaly-mediated SUSY breaking) resulting in the relatively narrow
bands.  {Thus, the annihilation cross-section predictions for gamma-ray
production from  higher energy ($\sim$100~GeV--TeV) candidates are well
constrained, with the particle-physics uncertainty contributing $\sim$ one
order of magnitude to the range of the predicted gamma-ray fluxes.} 

We elaborate further on the potential of $\gamma$-ray experiments to play a
pivotal role in identifying the dark matter particle and in particular, how a
next-generation $\gamma$-ray experiment can in fact provide information on the
actual formation of structure in the Universe.

\section{Dark Matter Annihilation into $\gamma$-rays, and the uncertainties
in the predicted flux}

For any of the scenarios that have been considered, the dark-matter particle
must be neutral and does not couple directly to photons, however most
annihilation channels ultimately lead to the production of photons through a
number of indirect processes.  While the total cross-section for gamma-ray
production is constrained by the measured relic abundance of dark matter, the
shape of the gamma-ray spectrum is sensitive to the details of the specific
particle-physics scenario.  Summarizing the previous discussion, dark matter
annihilation may yield photons in three ways: (1) by the direct annihilation
into a two-photon final state (or a $Z^0 \gamma$ or $H\gamma$ final state)
giving a nearly monoenergetic line,  (2) through the annihilation into an
intermediate state (e.g. a quark-antiquark pair), that subsequently decays and
hadronizes, yielding photons through the decay of neutral pions and giving rise
to a broad featureless continuum spectrum or (3) through
internal-bremsstrahlung into a three-particle state, e.g. $\chi\chi\rightarrow
W^+W^-\gamma$ yielding gamma-rays with a very hard spectrum and sharp cutoff.
The cross section for the direct annihilation into two photons, or a photon and
$Z^0$ are loop-suppressed and can be at least 2 orders of magnitude less than
the processes that lead to the continuum emission.  However, for some cases of
interest (e.g., a massive Higgsino) the annihilation line can be substantially
enhanced.  Also, in the next-to-minimal supersymmetric standard model (NMSSM)
with an extended Higgs sector, one-loop amplitudes for NMSSM neutralino pair
annihilation to two photons and two gluons, extra diagrams with a light CP-odd
Higgs boson exchange can strongly enhance the cross-section for the
annihilation line.  Such models have the added feature of providing a mechanism
for electroweak baryogenesis \cite{ferrer06}.  {{By combining Fermi
measurements of the continuum, with higher energy constraints from ground-based
ACT measurements, one can obtain constraints on the line to continuum ratio
that could provide an important means of discriminating between different
extensions to minimal supersymmetry or other dark matter scenarios}}

In general, the flux of $\gamma$-rays from a high-density annihilating
region can be written as 
\begin{equation} 
\frac{dN_\gamma}{dAdt} =L  {\cal P} 
\end{equation}
where,
\begin{equation} 
L  = \frac{1}{4 \pi} \int_{\rm LOS} \rho^2(r) dl 
\label{eq:ldef}
\end{equation}
contains the dependence to the
distribution of dark matter, and
\begin{equation} 
{ \cal P} = \int_{{E_{th}}}^{{M_\chi}}\sum_i \frac{
\langle \sigma v \rangle_i }{M_\chi^2} \frac{dN_{\gamma, i}}{dE} \, dE
\end{equation}
is the particle physics function that contains the detailed physical
properties of the dark matter particle. The sum over
the index $i$ represents the sum over the different photon production
mechanisms.  (In Eq.~\ref{eq:ldef}, $M_\chi$ is the neutralino mass,
$l$ is the line-of-sight distance
while $r$ is the radial distance from the center of the halo distribution.
Note that this definition of $L$ is similar to the definition of the $J$-factor
used elsewhere in the literature (e.g., \cite{bub98})  

Given the fact that supersymmetry has not been detected yet, the uncertainty in
the value of ${\cal P}$ is rather large. Sampling of the available
supersymmetric parameter space reveals that the uncertainty in cross sections
can be as large as 5 orders of magnitude if one covers the entire mass range
down that extends over several orders of magnitude (see
Fig.~\ref{fig:gondolosigma}), but collapses considerably for $M_\chi \gsim$100
GeV.  For supersymmetric dark matter, ${\cal P}$ can take a {\it maximum} value
of approximately ${\cal P} \approx 10^{-28} \, {\rm cm}^3 {\rm s}^{-1} {\rm
GeV^{-2}}$ when $M_\chi \approx 46 \, {\rm GeV}$, $\sigma v = 5 \times 10^{-26}
\, {\rm cm^3} {\rm s^{-1}}$ and $E_{\rm th}=5 \, {\rm GeV}$ (with a more
typical value of $\approx 2\times 10^{-26} \, {\rm cm}^3{\rm s}^{-1}$ at
energies between 100 GeV and 1 TeV . On the other hand, for a threshold energy
of $E_{\rm th}=50 \, {\rm GeV}$ and a particle mass of $M_\chi \approx 200
\,{\rm GeV}$, the value is ${\cal P} \approx 10^{-31} \, {\rm cm}^3 {\rm
s}^{-1} {\rm GeV^{-2}}$. 

It is important to emphasize that even though the actual value of ${\cal P}$
from supersymmetry can be orders of magnitude smaller, in theories with
universal extra dimensions, both the cross section into a photon final state
and the mass of the particle can actually be higher than this value.  

The quantity $L$, on the other hand,
contains all the information about the spatial distribution of
dark matter. Specifically, $L$ is proportional to the
line of sight (LOS) integration of the square of the dark matter density.
Dissipationless N-body simulations suggest the density profiles of dark
matter halos can be described by the functional form
\begin{equation}
\rho (\tilde{r}) = \frac{ \rho_s }{\tilde{r}^\gamma 
( 1 + \tilde{r} ) ^{\delta-\gamma}}
\label{eq:densityfunction}
\end{equation}
where $\tilde{r} = r / r_s$ (e.g., \cite{nfw97,moore99}). 

The quantities $\rho_s$ and $r_s$ are the characteristic density and radius
respectively, while $\gamma$ sets the inner, and $\delta$ the outer slope  of
the distribution. Recent simulations suggest that $\delta \approx 3 $, while
the value  of $\gamma$ has a range of values, roughly $0.7 \le \gamma \le 1.2$
down to $\sim 0.1 \%$ of the virial radius of the halo \cite{Netal,DZMSC}. A
change in the value of the inner slope $\gamma$ between the values of 0.7 and
1.2 for a fixed halo mass results in a change in the value of $L$ that is
roughly 6 times smaller or higher respectively \cite{SKBK06}.  The values of
$\rho_s$ and $r_s$ for a dark matter halo of a given mass are obtained if one
specifies the virial mass and concentration parameter. In general $\rho_s$ (or
the concentration parameter) depends solely on the redshift of collapse, while
$r_s$ depends on both the mass of the object as well as the redshift of
collapse.  In many previous studies the ``fiducial'' halo profile is that of
Navaro, Frenck and White (NFW; \cite{nfw97}) derived from an empirical fit to
the halo profile determined by N-body simulations and corresponding to
Eq.~\ref{eq:densityfunction} with $\delta=3$ and $\gamma=1$.

The main difficulty in estimating the value of $L$ for a dark matter halo is
due to the unknown density profiles in the regions from which the majority of
the annihilation flux is emitted.  Experimental data on the inner kiloparsec of
our Galactic (or extragalactic) halos is sparse and theoretical understanding
of these density profiles is limited by our lack of knowledge about the initial
violent relaxation in dark matter halos, and the complicated physics behind the
evolutionary compression of DM during the condensation of baryons in galactic
cores. Both processes still lack a complete theoretical understanding. The
uncertainty in the first is due to the unknown spectrum of density fluctuations
at small spatial scales and difficulties of predicting their evolution in high
resolution numerical simulations.  The uncertainty in the second is due to the
complexity of the gravitational interaction of the dark matter with the
dissipative baryonic matter on small scales and in regions of high density.
Experimentally, measurements of rotation curves and stellar velocity dispersion
are limited by finite angular resolution and geometric projection effects.
While progress is being made on both theoretical and experimental fronts, large
uncertainties remain.

\section{Targets for Gamma-Ray Detection}

The Galactic center has been considered the most promising target for the
detection of dark matter annihilation, with a flux more than an order of
magnitude larger than any potential galactic source (e.g., \cite{bub98}).  The
detection of $\gamma$-rays from the region of Galactic Center by the Whipple
and H.E.S.S. collaborations~\cite{kosack04, GCHESS2004} can, in principle,
include a contribution from annihilating dark matter~\cite{Horns2005}.  While
the flux and spectra of the Whipple and HESS detections are in agreement, the
Cangaroo-II group  reported the detection of high-energy gamma-ray emission
from the GC region \cite{tsuchiya04}, with a considerably softer spectrum that
now appears to be a transient effect (due to a variable source, or spurious
detection) in view of the latest, detailed HESS results. 

\begin{figure}[tb]
\includegraphics[width=0.46\textwidth]{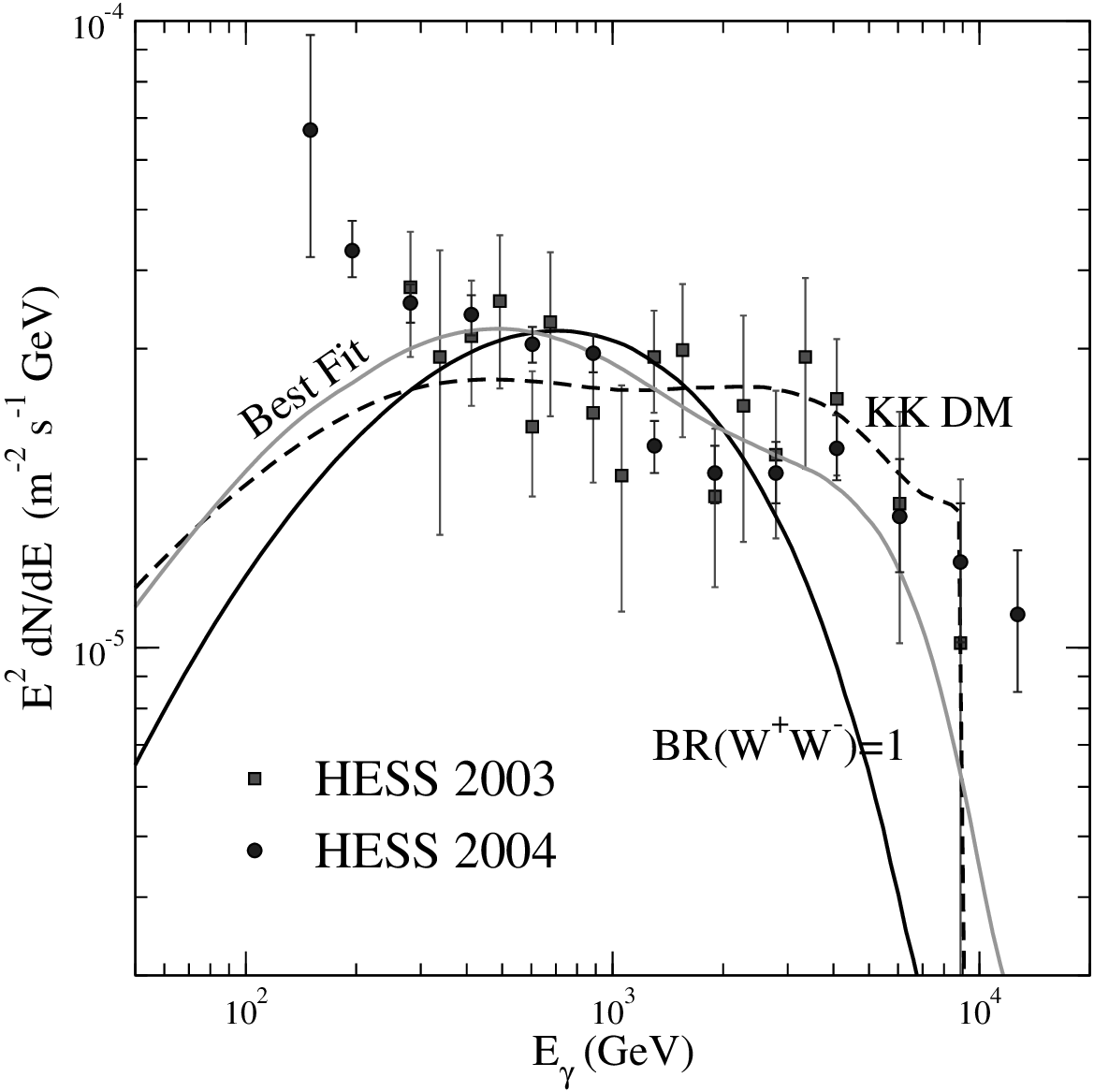}
\caption{
The HESS 2003 (grey squares) and HESS 2004 (filled circles) data on the flux of
GR from the GC, and the best fit to those data with a KK $B^{(1)}$
pair-annihilating lightest KK particle (dashed line), with a WIMP annihilating
into a $W^+W^-$ pair (black solid line), and with the best WIMP spectral
function fit (light grey line).}
\label{fig:dnde}
\end{figure}

\begin{figure}[tb]
\includegraphics[width=0.48\textwidth]{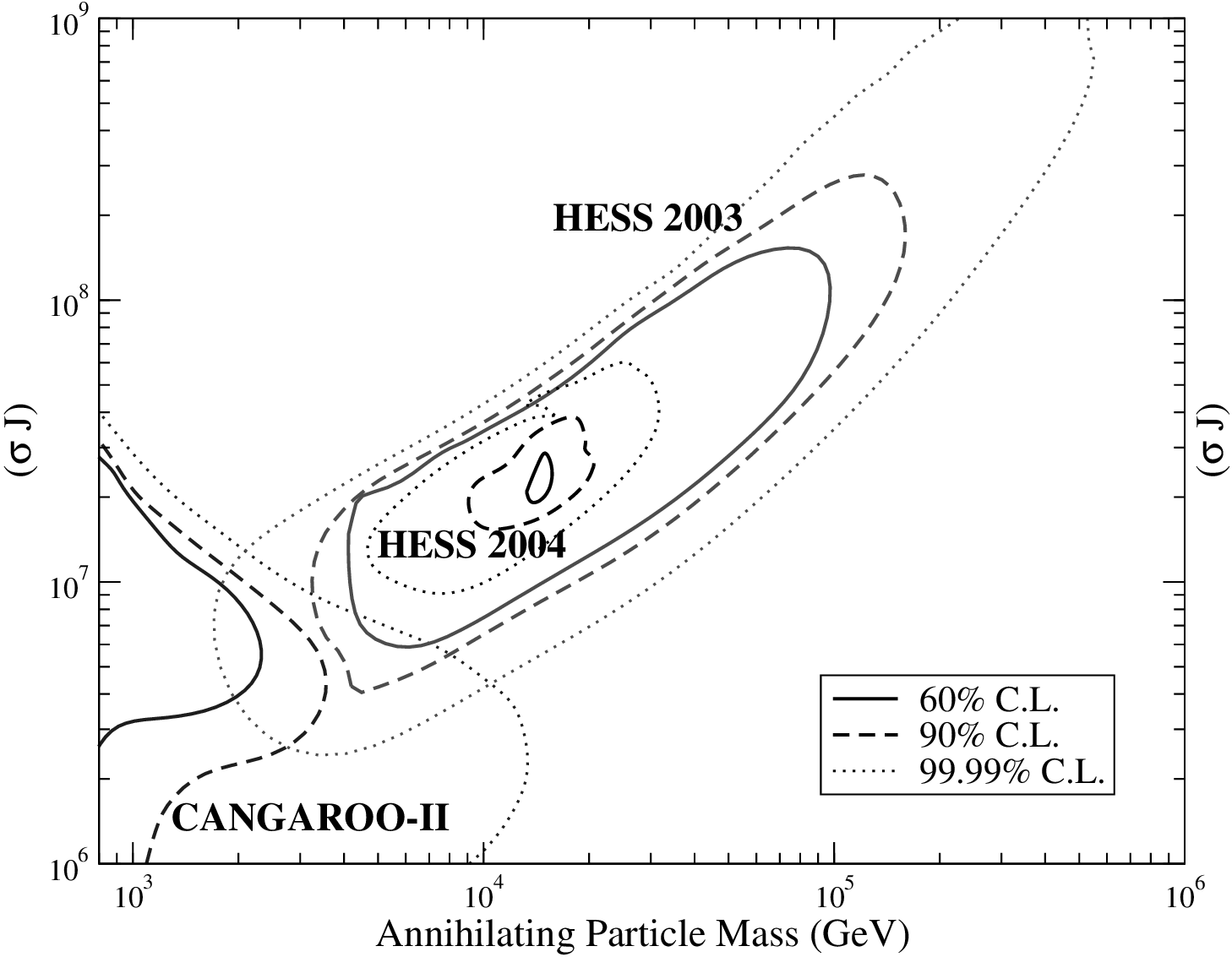}
\caption{
Iso-confidence-level contours of ``{\em best spectral functions}'' fits to the
Cangaroo-II and to the 2003 and 2004 HESS data, in the plane defined by the
annihilating particle mass and by the quantity $(\sigma J)$.
}
\label{fig:chi2_cntr}
\end{figure}

In Ref.~\cite{Profumo:2005xd} the possibility of interpreting the GR data from
the GC in terms of WIMP pair annihilations was analyzed in full generality.
Examples of fits to the HESS data with a Kaluza-Klein (KK) $B^{(1)}$ DM
particle, with WIMPs annihilating into $W^+W^-$ in 100\% of the cases and with
the best possible combination of final states, namely $\sim30$\% into $b\bar b$
and $\sim70$\% into $\tau^+\tau^-$ are shown in fig.~\ref{fig:dnde}. Those
options give a $\chi^2$ per degree of freedom of around 1.8, 2.7 and 1: only
the best-fit model is found to be statistically viable.

Using the Galactic-center data and assuming that the observed gamma-ray
emission arises from dark-matter annihilation, Profumo \cite{Profumo:2005xd}
derived confidence intervals for the product of the total annihilation
cross-section $\sigma$ and the $J$-factor (characterizing the astrophysical
uncertainty from the halo density profile) versus the neutralino mass $m_\chi$.
Iso-confidence-level contours in the $(m_{\chi},(\sigma {J}))$ plane are shown
in fig.~\ref{fig:chi2_cntr}. From the figure, it is clear that a dark-matter
origin for the emission requires a DM mass range between 10-20 TeV. Further, a
value of $(\sigma{J})\approx 10^7$ implies either a very large astrophysical
boost factor ($\approx 10^3$ larger than what expected for a NFW DM profile),
or a similar enhancement in the CDM relic abundance compared with the
expectations for thermal freeze-out

Ref.~\cite{Profumo:2005xd} showed that some supersymmetric models can
accommodate large enough pair annihilation cross sections and masses to both
give a good fit to the HESS data and thermally produce the right DM abundance
even though, from a particle physics point of view, these are not the most
natural models.  An example is a minimal anomaly-mediated SUSY breaking
scenario with non-universal Higgs masses.  For some choices of model
parameters, such a dark matter particle could even be directly detected at
ton-sized direct detection experiments, even though the lightest neutralino
mass is in the several TeV range \cite{Profumo:2005xd}.

However, the interpretation is particularly complicated since the center of our
own Milky Way galaxy has a relatively low mass-to-light ratio and is dominated
by matter in the form of a central massive black hole and a number of other
young massive stars, supernova remnants and compact stellar remnants.
Moreover, the lack of any feature in the power-law spectrum measured b HESS,
and the extent of this spectrum up to energies above 10 TeV makes a dark-matter
interpretation difficult.

A way of dealing with this background is to exclude the galactic center source
seen by HESS, and instead look at an annulus about the Galactic center position
\cite{Stoehr:2003hf, serpico08}.  Even though the background grows in
proportion to the solid angle of the annular region (and the sensitivity
degrades as the square-root of this solid angle) for a sufficiently shallow
halo profile, the signal-to-noise ratio for detection continues to grow out to
large angles.  Moreover, any component of diffuse contaminating background
falls off more steeply as a function of latitude than the annihilation of the
smooth component of the dark matter halo. This result may even be enhanced by
the presence of other bound high density structures within the inner parts of
the Milky Way \cite{Diemand:2006ik}. 

\begin{figure*}[tb]
\begin{center}
\includegraphics[height=10cm]{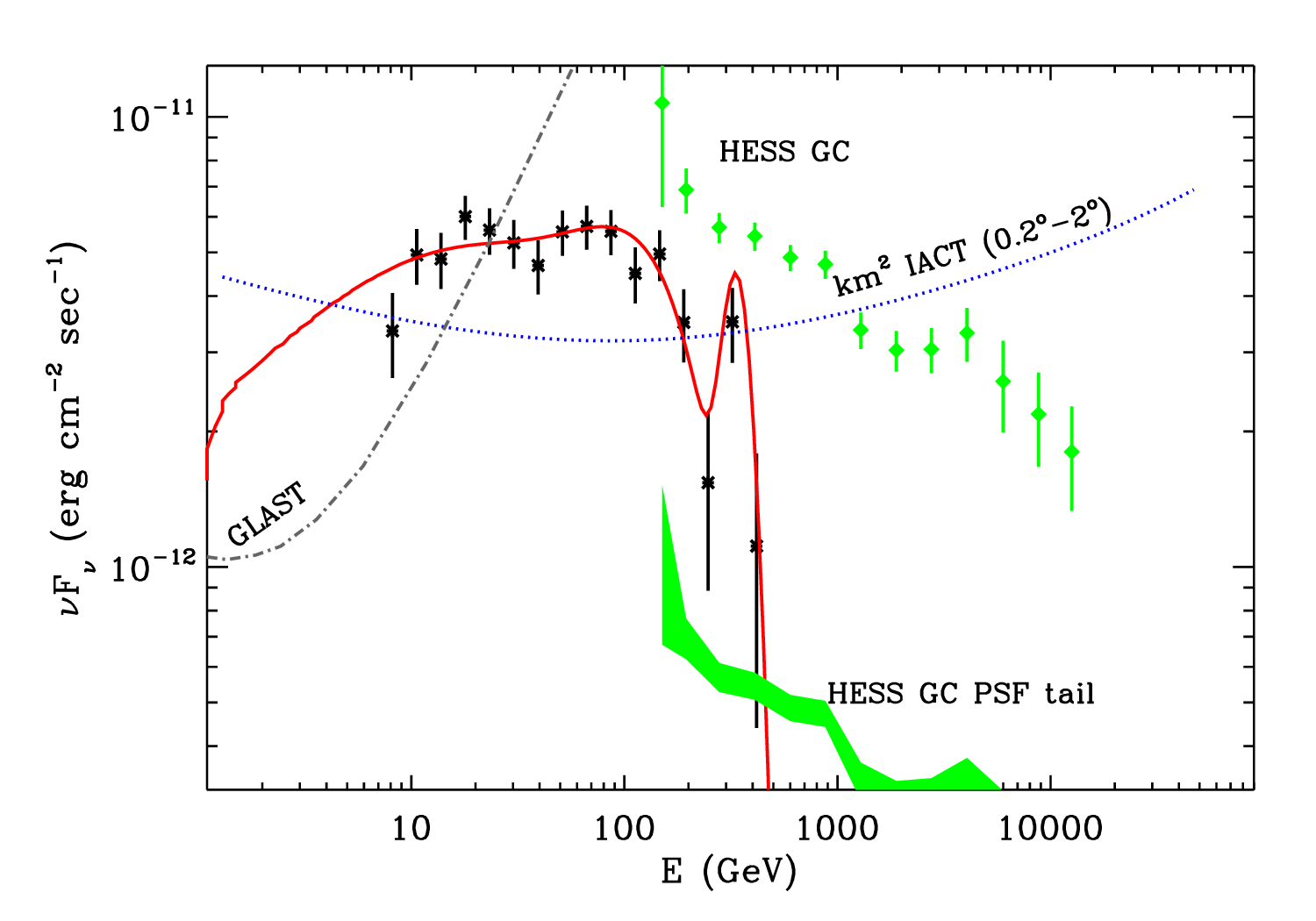}
\end{center}
\caption{ Gamma-ray spectrum from dark matter annihilation in an annulus
between 0.2$^\circ$ and 2$^\circ$ about the Galactic center assuming an NFW
halo with a central density of $\rho_s = 5.4\times 10^6\, M_\odot/{\rm kpc}^3$
and a scale radius of $r_s = 21.7\,{\rm kpc}$.   We show the HESS spectrum of
the point source near the GC, and 10\% of this value assumed to bleed into the
annulus from the tails of the gamma-ray point-spread-function.  Here we assume
a 200~hour exposure of a a km$^2$ IACT instrument.  The reduced sensitivity,
compared with that for a point source, comes from integrating the hadronic,
electron, and diffuse gamma-ray background over the relatively large solid
angle of the annulus.  }
\label{fig:gcannulus}
\end{figure*}

We make a conservative estimate of the signal from an annulus centered on the
galactic center.  For this calculation, we assume that the Milky Way halo has a
profile as given by Navaro, Frenck and White \cite{nfw97} (NFW profile) with a
scale radius of $r_s = 21.7$~kpc and a central density of $\rho_s=5.38\,
M_\odot\, {\rm kpc}^{-3}$ from Fornengo et al. \cite{fornengo04}.  To be
somewhat more conservative, in light of more recent N-body simulations that
show a flattening of the inner halo profile, we assume a 10~pc constant density
core.  The minimum angle for the annular region is set by the assumed PSF for a
future instrument.  We assume that the flux from the point source at the GC (or
from the diluted contribution from the galactic ridge emission) will fall below
10\% of the GC value, 0.2 deg from the position of Sgr~A*.  The optimum angular
radius for the outer bound on the annulus is 12~deg (see \cite{serpico08} for
details), somewhat beyond the largest field of view envisioned for a future
imaging ACT (with a more realistic value of 6-8~deg).  As shown in
Fig.~\ref{fig:gcannulus}, Fermi might also have adequate sensitivity and
angular resolution to detect the continuum emission and separate this from the
other point sources.  If the neutralino mass is large enough (above several
TeV) and one chooses favorable parameters for the annihilation cross-section
and density, EAS detectors have the large field-of-view required to observe
such extended sources as well as other regions of emission along the galactic
plane.  However, these detectors lack the good angular and energy resolution to
separate this emission from other point sources and would require follow-up
observations by more sensitive instruments such as imaging ACT arrays.  For the
IACT sensitivity, we assume that we have an instrument with effective area of
1~km$^2$, an exposure of 200~hrs, and that the background comes from cosmic-ray
electrons, cosmic-ray atmospheric showers, and diffuse gamma-rays following the
method given in Ref.~\cite{bub98}.  For the diffuse gamma-ray spectrum, we take
the EGRET diffuse flux, and assume that it continues with a relatively hard
$\sim E^{-2.5}$ spectrum up to TeV energies.  We also assume that the largest
practical angular radius of the annular region is 2~deg, a reasonable value for
a moderately wide-field-of view future instrument.  The simulated spectrum is
calculated for a typical annihilation cross section of $\langle \sigma v\rangle
= 2\times 10^{-26}{\rm cm}^3{\rm s}^{-1}$ and for an arbitrary set of branching
ratios corresponding to 50\% $\tau\bar{\tau}$, 50\% $b\bar{b}$ and a
line-to-continuum ratio of $6\times 10^{-3}$.  Assuming a 15\% energy
resolution, we obtain the simulated spectrum shown in Fig.~\ref{fig:gcannulus}.
This demonstrates that a future instrument could observe a spectral signature
of dark matter annihilation in the region around the GC, above the residual
astrophysical backgrounds.  To search for gamma-ray emission from dark-matter
annihilation in the Galactic center region, the requirements for the future
instrument include: a large effective area ($\sim$1~km$^2$), a moderately large
field of view ($\gsim 7^\circ$ diameter), a good energy resolution ($\lsim
15$\%), a low energy threshold ($\lsim 50$~GeV), excellent angular resolution
to exclude contributions from astrophysical point-sources ($<\sim 0.1^\circ$)
and a location at low geographic latitude (preferably in the southern
hemisphere) for small-zenith-angle low-threshold measurements of the GC region.

However, given the large backgrounds in our own galaxy, the observation of a
wider class of astrophysical targets is desirable.  A future km$^2$ ACT array
should, for the first time, have the sensitivity required to detect
extragalactic sources such as Dwarf galaxies, without resorting to very
optimistic assumptions about the halo distribution.  The VERITAS collaboration
previously undertook such an observing program with the Whipple 10m telescope
and reported upper limits for several extragalactic targets (M33, Ursa Minor \&
Draco dwarf galaxies, M15)~\cite{Vassiliev2003,LeBohec2003,woodetal07}.   The
HESS group published limits on the Sagittarius dwarf galaxy and the resulting
constraints on the halo models \cite{moulin07}.  However, more sensitivity is
required to detect a more generic annihilation flux from such sources. 

\begin{figure*}[tb]
\begin{center}
\includegraphics[height=10cm]{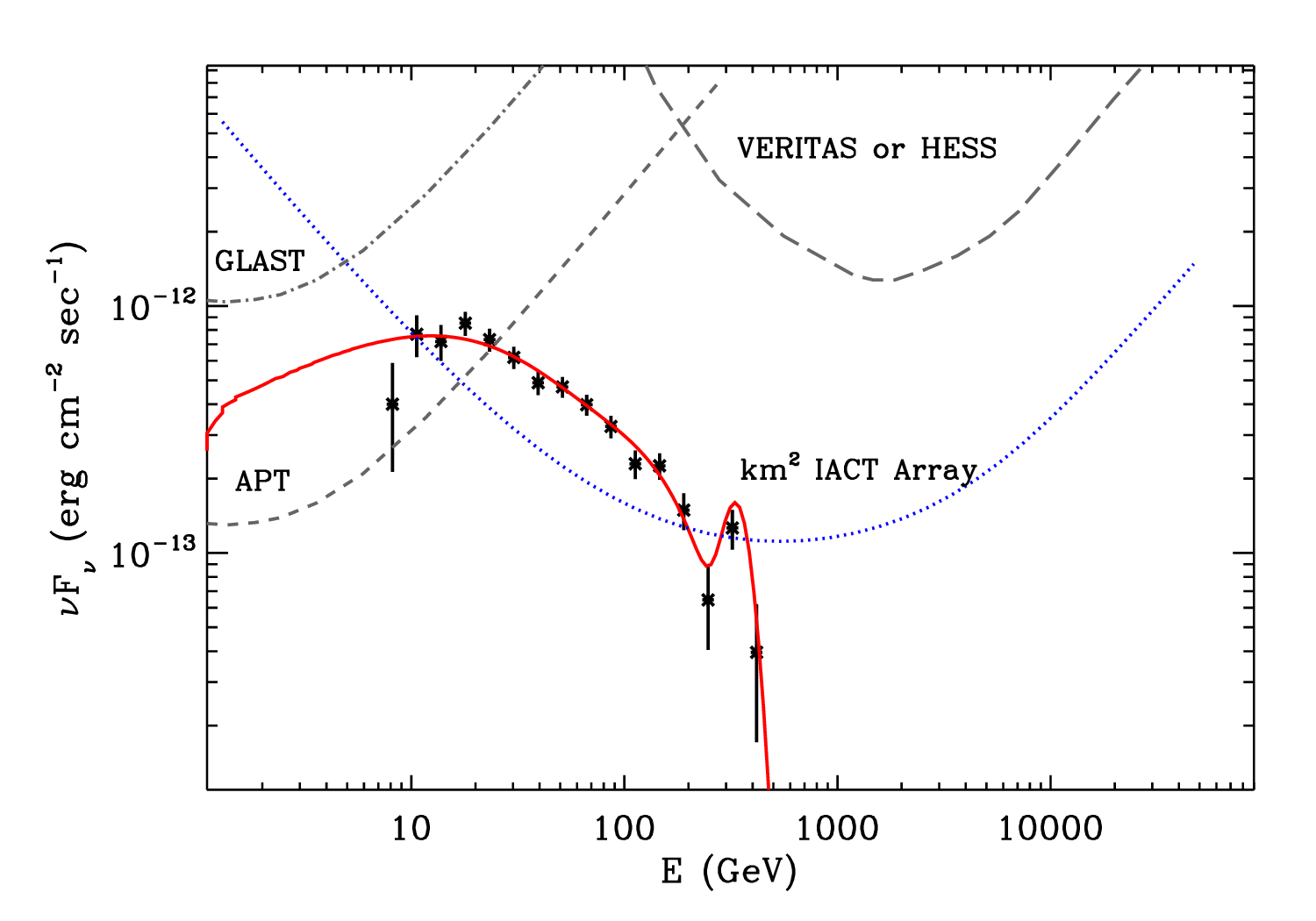}
\end{center}
\caption{ Predicted gamma-ray signal from the dwarf spheroidal galaxy Ursa
Minor for neutralino mass of 330 GeV, branching into $\tau^+\tau-$ 20\% of the
time, and into $b\bar{b}$ 80\% of the time and with a line to continuum ratio
of 2$\times 10^{-3}$.  We assume a typical annihilation cross-section of
$2\times 10^{-26}{\rm cm}^3{\rm s}^{-1}$ the halo values from Strigari et al.
\cite{Strigari:2007at} with $r_s = 0.86\,{\rm kpc}$ and central density $\rho_s
= 7.9\times 10^7\, M_\odot/{\rm kpc}^3$.  We also assume a modest boost factor
of $b=3$ from halo substructure.  We assume an ideal instrument with an
effective area of 1~km$^2$ and sensitivity limited only by the electron
background, diffuse gamma-ray background (assuming an $\sim E^{-2.5}$ spectrum
connecting to the EGRET points) and cosmic-ray background (10 times lower than
current instruments).  For this idealized IACT array, we do not include the
effect of a threshold due to night-sky-background, and assume an energy
resolution of 15\%.  The data points are simulated given the signal-to-noise
expected for the theoretical model compared with our anticipated instrument
sensitivity.  }
\label{fig:ursaminor}
\end{figure*}

\subsection{Dwarf Spheroidals} Dwarf spheroidal (dSph) systems are ideal
dark matter laboratories because astrophysical  backgrounds and baryon-dark
matter interactions are expected not to play a major role in the distribution
of dark matter. Furthermore, the mass--to--light ratio in dSphs can be very
large, up to a few hundred, showing that they are largely dark-matter dominated
systems.  Numerous theoretical studies point to the potential for detecting
dark matter annihilation in dwarf spheroidal galaxies or galaxies in the local
group based on rough assumptions of the distribution of dark matter
\cite{Profumo:2005xd,Betal02,Tyler02, BH06,PB04}.  However, with the advent of
more data on the stellar content of dSphs, it has recently been possible to
perform a likelihood analysis on the potential dark matter profiles that these
systems could posses.  Under the assumption that dSphs are in equilibrium, the
radial component of the stellar velocity dispersion is linked to the
gravitational potential of the system through the Jeans equation. This approach
(utilized in \cite{EFS04, SKBK06,Strigari:2007at}) has the significant
advantage that observational data dictate the distribution of dark matter with
a minimum number of theoretical assumptions.  The main results of these studies
are that dSphs are very good systems for the search for dark matter
annihilation, because most of the uncertainties in the distribution of dark
matter can be well quantified and understood.  In addition, dSphs are expected
to be relatively free of intrinsic $\gamma$-ray emission from other
astrophysical sources, thus eliminating contaminating background that may
hinder the interpretation of any observation.  Assuming a scenario for
supersymmetric dark matter where $M_\chi = 200$\,GeV, $E_{\rm th}=50$\,GeV and
${\cal P} \approx 10^{-31} \, {\rm cm}^3 {\rm s}^{-1} {\rm GeV^{-2}}$, the
maximum expected fluxes from 9 dSphs studied in \cite{SKBK06,Strigari:2007at}
can be as large as $10^{-12}$ photons cm$^{-2}$ s$^{-1}$ (for Willman 1).
Observing $\gamma$-rays from dark matter annihilation in dwarf spheroidals is
of fundamental importance for 2 reasons: First and foremost, these observations
can lead to an identification of the dark matter, especially if line emission
or other distinct features in the continuum are detected and second, they will
provide information on the actual spatial distribution of dark matter halos in
these important objects.  If there is a weakly interacting thermal relic, then
$\gamma$-ray telescopes can tell us something about non-linear structure
formation, a task unattainable by any other experimental methods.

Fig.~\ref{fig:ursaminor} shows an example of one possible spectrum that might
be measured for Ursa Minor given conservative assumptions including: a typical
annihilation cross-section, a halo distribution constrained by stellar velocity
measurements ( from Strigari et al. \cite{Strigari:2007at}) and a modest boost
factor of $b=3$ at the low end of the expected range for such halos.  This
prediction demonstrates that detection from Dwarf galaxies is most likely out
of reach of the current generation of IACT experiments (HESS and VERITAS) or
proposed EAS experiments, but may be within reach of a future km$^2$IACT
instrument, if the point-source sensitivity is improved by an order of
magnitude, the energy resolution is good enough to resolve the spectral
features (better than 15\%) and the energy threshold can be pushed well below
100~GeV.    

With the advent of the Sloan Digital Sky Survey (SDSS), the number of known
dSph satellites of the local group has roughly doubled during the last decade
\cite{belokurov}.  Since the survey is concentrated around the north Galactic
pole, it is quite likely that there are many more dSph satellites waiting to be
discovered. For an isotropic distribution, and assuming that SDDS has found all
the satellites in its field of view, we would expect $\sim$ 50 dwarfs in all.
Since simulation data suggests that dwarf satellites lie preferentially along
the major axis of the host galaxy, the number of Milky-Way dwarf satellites
could be well above this estimate.  With more dwarf galaxies, and increasingly
detailed studies of stellar velocities in these objects, this class of sources
holds great promise for constraints on dark matter halos and indirect detection
of dark matter.  Since many of these discoveries are very new, detailed
astronomical measurements are still required to resolve the role of dark matter
in individual sources.  For example, for the new object Willman I, some have
argued that this is a globular cluster while others have made the case that
despite it's relatively small mass, this is a dark-matter dominated object and
not a globular cluster \cite{martin07}.  Other studies challenge the inferences
about the dark matter dominance in dSphs attributing the rise in rotation
velocities in the outer parts of dSphs to tidal effects rather than the
gravitational potential \cite{metz}.  Future progress in this blossoming area
of astronomy could provide important additional guidance for a more focused
survey on the most promising sources using pointed observations with very deep
exposures. 

\begin{figure}[tb]
\includegraphics[width=0.48\textwidth]{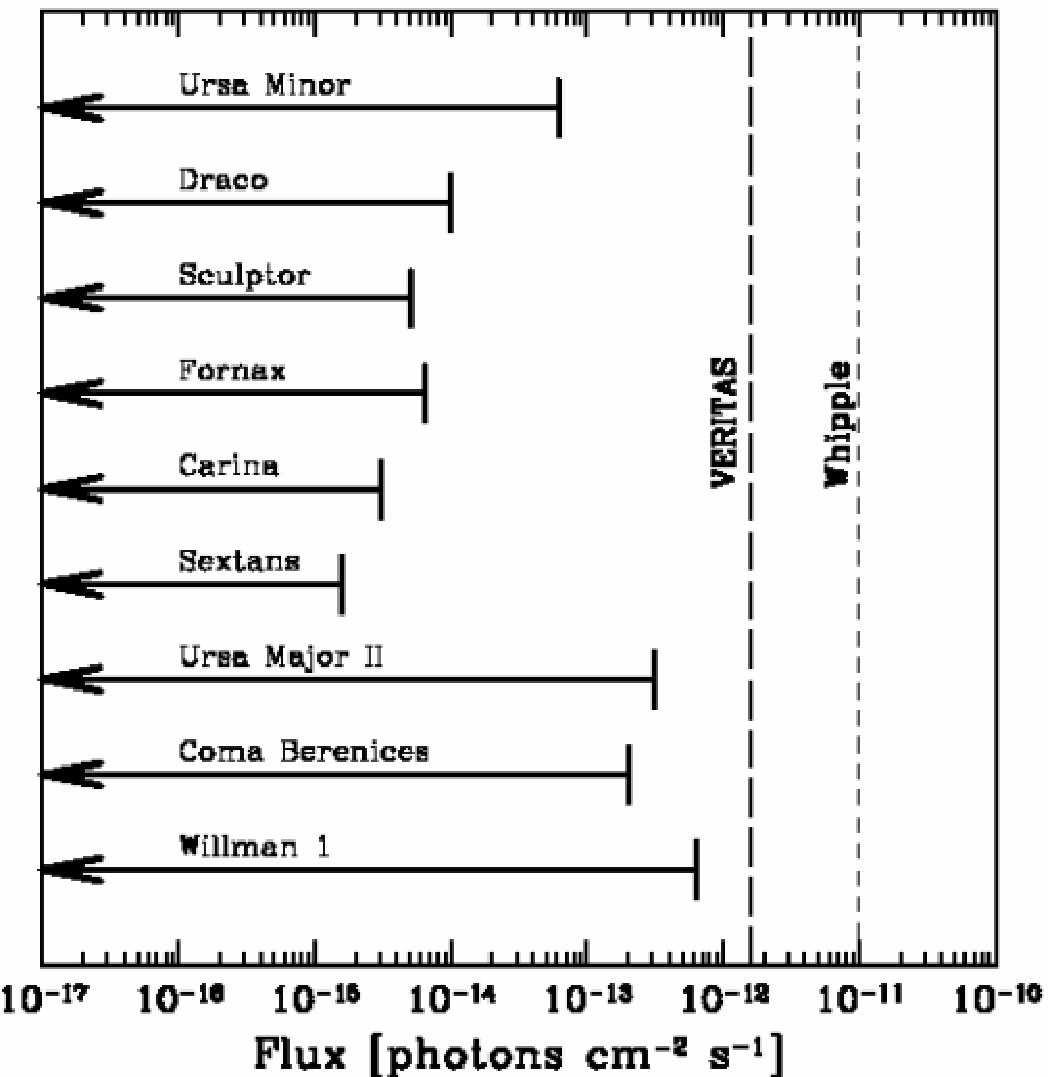}
\caption{Prospects for detecting the most prominent Dwarf-galaxy targets for
dark matter annihilation. Upper-limit bars show the range of theoretical
predictions \cite{koushiappas_santafe06} with fluxes dropping below the level
of detectability as one traverses the full range of parameter space including
the neutralino mass, cross-section and halo distribution. The plot includes
dark-matter dominated dwarf spheroidal systems in the Milky Way halo, including
promising sources located at high galactic latitude and with virtually no known
intrinsic $\gamma$ ray emission from astrophysical sources. The thin-dashed
line represents the sensitivity of Whipple, while the long-dashed line depicts
the sensitivity of VERITAS.}
\label{fig:dSphs}
\end{figure}

\subsection{Local group galaxies} Local group galaxies offer attractive
targets for the search of $\gamma$-rays form dark matter annihilation for many
of the same reasons dSph galaxies do: they are relatively small systems, with
relatively high mass-to-light ratios (except M31). Relative to dSphs, the
influence of baryons in the central regions is higher, especially if a black
hole is present (such as M32). Nevertheless, their relative proximity and size
make them viable targets that should be explored. Recently, Wood et al. (2007)
\cite{woodetal07} used the Whipple 10m telescope and placed bounds on the
annihilation cross section of neutralinos assuming a distribution of dark
matter in the halos of M32 and M33 that resembles dark matter halos seen in
N-body simulations.  While these observations with Whipple and now with VERITAS
and HESS provide interesting limits on some of the more extreme astrophysical
or particle physics scenario, more sensitive observations are needed if one
makes more conservative estimates.  Even with an order of magnitude increase in
sensitivity over the current generation experiments, it is still possible that
Dwarf or local-group galaxies will evade detection with the next generation
detector without some enhancement in the central halo (e.g. a cusp steepened by
the stellar population or a large boost factor).  Given this uncertainty {the
best strategy for detecting dark matter from Dwarf galaxies, or local group
galaxies is to observe an ensemble of sources, taking advantage of the
source-to-source variance in the halo profile until better constraints are
available from new astronomical measurements (e.g., stellar velocity dispersion
or rotation curves).}

\subsection{Detecting the Milky Way\\Substructure}

A generic prediction of the hierarchical structure formation scenario in cold
dark matter (CDM) cosmologies is the presence of rich substructure; bound dark
matter halos within larger, host halos.  Small dark matter halos form earlier,
and therefore have higher characteristic densities.  This makes some of these
subhalos able to withstand tidal disruption as they sink in the potential well
of their host halo due to dynamical friction. Unfortunately, even though this
is a natural outcome of CDM, there is no clear explanation as to why the Milky
Way appears to contain a factor of 10-100 {\it fewer} subhalos than it should,
based on CDM predictions \cite{Klypin:1999uc,Moore:1999wf}.  Several solutions
to this problem have been suggested, such as changing the properties of the
dark matter particle (e.g.,
\cite{1992ApJ...398...43C,Spergel:1999mh,Kaplinghat:2000vt}), modifying the
spectrum of density fluctuations that seed structure growth (e.g.,
\cite{Kamionkowski:1999vp, Zentner:2002xt}), or invoking astrophysical feedback
processes that prevent baryonic infall and cooling (e.g.,
\cite{1986ApJ...303...39D, 1999ApJ...523...54B,2001ApJ...548...33B}).  The most
direct experimental way to probe the presence of otherwise dark substructure in
the Milky Way is through $\gamma$-ray observations.  Theoretical studies
\cite{Koushiappas:2003bn}, as well as numerical simulations of a Milky Way-size
halo \cite{Diemand:2006ik}, predict that given the probability of an otherwise
completely dark subhalo nearby, the expected flux in $\gamma$-rays can be as
large as $\sim 10^{-13}$ cm$^{-2}$ s$^{-1}$. 

\subsection{Detecting Microhalos}

The {\it smallest} dark matter halos formed are set by the RMS dark matter
particle velocities at kinetic decoupling, the energy scale at which
momentum--changing  interactions cease to be effective
\cite{Setal99,Hetal01,Chenetal01,Berezinskyetal03,Green04,
Green05,LZ05,Bertch}. For supersymmetric dark matter this cutoff scale fives a
mass range for {\it microhalos} of around $10^{-13} \le [M/M_\odot] \le
10^{-2}$, depending on the value of the  kinetic decoupling temperature which
is set by the supersymmetric parameters.  While the survival of microhalos in
the Solar neighborhood is still under debate, there are indications that some
fraction ($\sim 20\%$) may still be present. In this case, microhalos could
even be detected via the proper motion of their $\gamma$-ray  signal
\cite{Mooreetal06,K06}. Microhalos that exhibit proper motion must be close
enough that their proper motion is above a detection threshold set by the
angular resolution and length of time over which the source can be
monitored(given by the lifetime of the observatory).  Microhalos must be
abundant enough so that at least one is within the volume set by this proper
motion requirement. The expected flux from a microhalo that may exhibit
detectable proper motion \cite{K06} is $\sim 10^{-15}$ cm$^{-2}$ s$^{-1}$. 
Such objects are most likely to be detected by very wide-field instruments 
like Fermi.  Follow-up measurements with IACT arrays would be required to
determine the characteristics of the spectrum and angular extent of these
sources at higher energies.

\subsection{Spikes around Supermassive and Intermediate-Mass Black Holes}
There are other potential dark matter sources in our own Galaxy that may be
formed by a gravitational interplay of dark halos and baryonic matter.  In
particular, it is possible that a number of intermediate-mass black holes
(IMBHs) with cuspy halos, might exist in our own galaxy.  The effect of the
formation of a central object on the surrounding distribution of matter has
been investigated in
Refs.~\cite{peebles:1972,young:1980,Ipser:1987ru,Quinlan:1995} and for the
first time in the framework of DM annihilations in Ref.~\cite{Gondolo:1999ef}.
It was shown that the {\it adiabatic} growth of a massive object at the center
of a power-law distribution of DM, with index $\gamma$, induces a
redistribution of matter into a new power-law (dubbed ``spike'') with index
\begin{equation} \gamma_{sp} = (9-2\gamma)/(4-\gamma) \;\; .  \end{equation}
This formula is valid over a region of size $R_{sp} \approx 0.2 \, r_{BH}$,
where $r_{BH}$ is the radius of gravitational influence of the black hole,
defined implicitly as $M(<r_{BH})=M_{BH}$, where $M(<r)$ denotes the mass of
the DM distribution within a sphere of radius $r$, and where $M_{BH}$ is the
mass of the Black Hole~\cite{Merritt:2003qc}.  The process of adiabatic growth
is, in particular, valid for the SMBH at the galactic center.  A critical
assessment of the formation {\it and survival} of the central spike, over
cosmological timescales, is presented in
Refs.~\cite{Bertone:2005hw,Bertone:2005xv} and references therein.  Adiabatic
spikes are rather fragile structures, that require fine-tuned conditions to
form at the center of galactic halos~\cite{Ullio:2001fb}, and that can be
easily destroyed by dynamical processes such as major
mergers~\cite{Merritt:2002vj} and gravitational scattering off
stars~\cite{Merritt:2003eu,Bertone:2005hw}.

\begin{figure}[tb]
\label{fig:imbhs}
\includegraphics[width=0.48\textwidth]{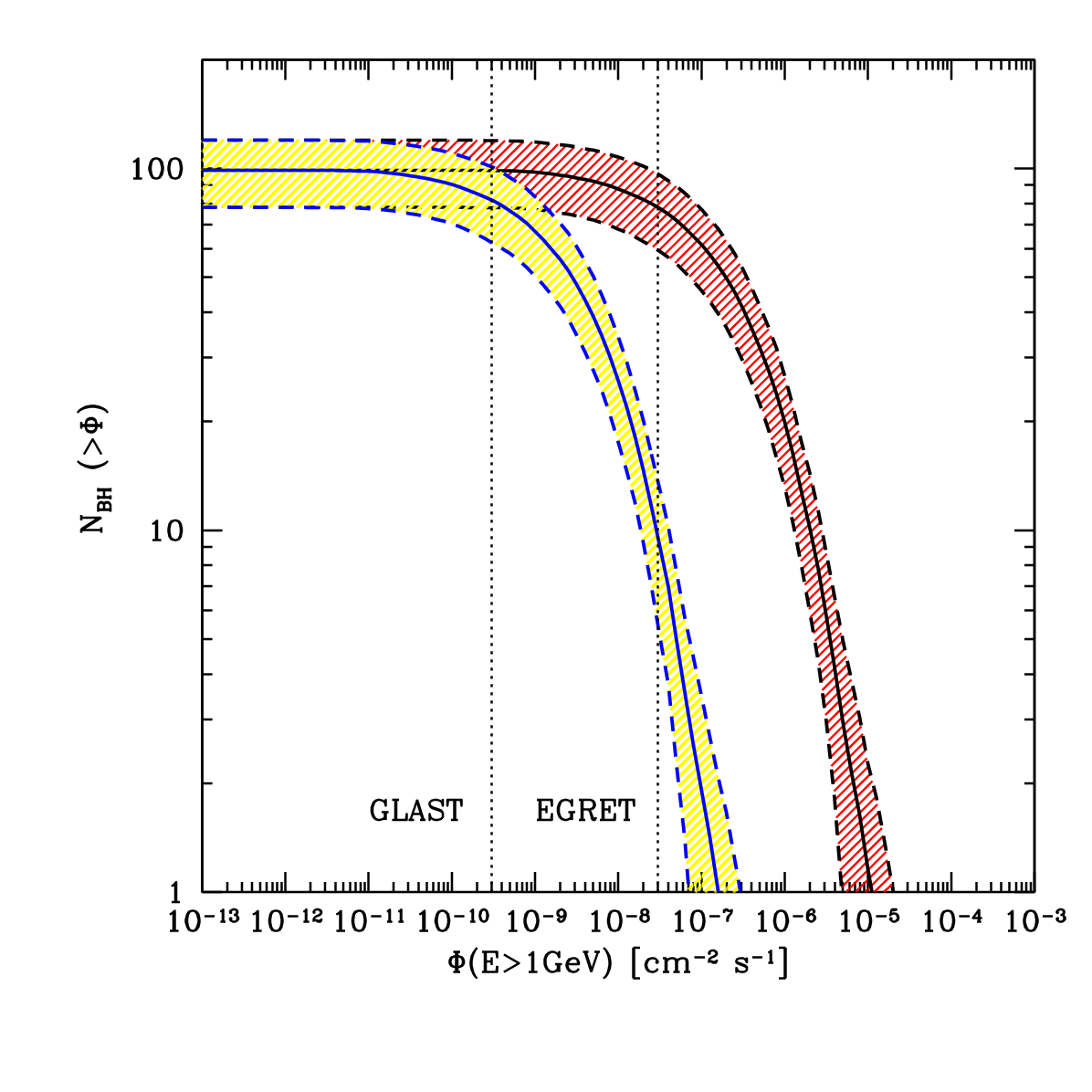}
\caption{IMBHs integrated luminosity function,
i.e. number of IMBHs that can be detected
from experiments with point source sensitivity $\Phi$ (above 1 GeV),
as a function of $\Phi$. We show for comparison the 5$\sigma$ point
source sensitivity above $1$~GeV of EGRET and Fermi (GLAST) in 1 year. From
Ref.~\cite{Bertone:2005xz}.}
\end{figure}

However Intermediate Mass BHs, with mass $10^{2} < M/{\rm M_{\odot}} < 10^{6}$,
are not affected by these destructive processes.  Scenarios that seek to
explain the observed population and evolutionary history of
supermassive-black-holes actually result in the prediction of a large
population of wandering IMBHs, with a number in our own Galaxy. They may form
in rare, overdense regions at high redshift, $z \sim 20$, as remnants of
Population III stars, and have a characteristic mass-scale of a few $10^{2} \,
{\rm M_{\odot}}$ \cite{Madau:2001,Bertone:2005xz,
Zhao:2005zr,islamc:2004,islamb:2004}.  Alternatively, IMBHs may form directly
out of cold gas in early-forming halos and are typified by a larger mass scale
of order $10^{5} \, {\rm M_{\odot}}$~\cite{Koushiappas:2003zn}. We show in
Fig.~\ref{fig:imbhs} the number of objects that can be detected as a function
of the detector sensitivity.  {The spiky halos around galactic
intermediate-mass black holes could provide a large enhancement in the
gamma-ray signal that could be effectively detected by all-sky low-threshold
instruments such as Fermi then followed-up by ground-based measurements.}  Over
most of the allowed parameter space, Fermi would detect the onset of the
continuum spectrum but would lack the sensitivity to measure the detailed
spectral shape above hundreds of GeV.  Ground-based measurements with good
point-source sensitivity, and good energy resolution (10-15\%) would be
necessary to follow-up these detections to measure the spectral cutoff and
other features of the annihilation spectrum needed to clearly identify a
dark-matter origin for the gamma-ray signal. 

High energy gamma-ray astronomy can also indirectly provide information about
the formation history of IMBHs through a very different avenue, i.e., infrared
absorption measurements of gamma-rays from distant AGN.  For example, the early
population-III stars that may seed the growth of IMBHs are likely to be massive
(100 $M_\odot$) stars that form in dark matter clumps of mass $\sim 10^6
M_\odot$.  These short lived stars would result in a large contribution to the
total amount of visible and UV light in the early (large-redshift) universe,
that contribute to the present-day diffuse infrared background.  Present
observations by Whipple, HEGRA, MAGIC and HESS already provide constraints on
the contribution from population-III stars.  {Gamma-ray astronomy has the
unique potential to provide important constraints on the history of structure
formation in the universe through observations of the annihilation signal from
dark-matter halos on a range of mass scales (including IMBH halos) in addition
to probing the history of star formation through measurements of the diffuse
infrared background radiation.}

\subsection{Globular clusters} Globular clusters are relatively low
mass-to-light ratio bound systems in the Milky Way that are dominated by a
dense stellar core. The presence of dark matter in the core of a collapsed
globular cluster is questionable because it is expected that 2-body stellar
interactions will deplete dark matter from the region. On the other hand, if
there is any dark matter left-over from the core-collapse relaxation process,
it is possible that the dense stellar core would adiabatically steepen the
distribution of dark matter, thus making some dense globular clusters potential
targets for dark matter detection.  Wood et al. (2007)  \cite{woodetal07}
observed the relatively close M15 globular cluster with the Whipple 10m
telescope, and placed upper bounds on the cross section for dark matter
annihilation.

\section{Complementarity of $\gamma$-Ray Searches with Other Methods for Dark Matter Searches}

Both Fermi and the LHC are expected to become operational in 2008.   What
guidance will these instruments provide for a future ground-based experiment?
The ATLAS and CMS experiments at the Large Hadron Collider (LHC) are designed
to directly discover new supersymmetric particles in the range of a few $\sim
100$ GeV/c$^{2}$ and will start collecting data in the very near future.  The
LHC alone will not, under even the most optimistic circumstances, provide all
of the answers about the nature of dark matter.  {In general, a combination of
laboratory (LHC, ILC) detection and astrophysical observations or direct
detection experiments will be required to pin down all of the supersymmetric
parameters and to make the complete case that a new particle observed in the
laboratory really constitutes the dark matter.} Due to the fact that the
continuum gamma-ray signal depends directly on the total annihilation
cross-section, there are relatively tight constraints on the gamma-ray
production cross-section from the cosmological constraints on the relic
abundance.  For direct detection, on the other hand, the nuclear recoil
cross-section is only indirectly related to the total annihilation
cross-section and thus there are a number of perfectly viable model parameters
that fall many orders of magnitude below any direct detection experiment that
may be built in the foreseeable future.  {Thus gamma-ray astronomy is unique in
that the detection cross-section is closely related to the total annihilation
cross section that determines the relic abundance}.  A given theoretical
scenario of SUSY breaking at low energies, e.g. mSUGRA, SplitSUSY,
non-universal SUGRA, MSSM-25, AMSB, etc., reduces the available parameter phase
space. Therefore, it is natural to expect that, for some set of the parameters,
the neutralino might be detected by all experimental techniques, while in other
cases only a single method has sufficient sensitivity to make a
detection~\cite{JE2004}. Only a combination of accelerator, direct, and
indirect searches would cover the supersymmetric parameter space~\cite{BG2002}.
For example, the mass range of neutralinos  in the MSSM is currently
constrained by accelerator searches to be above a few
GeV~\cite{Bottino2003,Bottino2004} and by the unitarity limit on the thermal
relic to be below $\sim 100$ TeV~\cite{GK1990} (a narrower region would result
if specific theoretical assumptions are made, e.g. mSUGRA).

For the LHC to see the lightest stable SUSY particle, it must first produce a
gluino from which the neutralino is produced.  This limits the reach of the LHC
up to neutralino masses of $m_\chi \approx 300$GeV, well below the upper end of
the allowed mass range.  Direct detection of WIMP-nucleon recoil is most
sensitive in the $60$ to $600$ GeV regime. Indirect observations of
self-annihilating neutralinos through $\gamma -$rays with energies lower than
$\sim 100$ GeV will best be accomplished by Fermi, while VERITAS and the other
ground-based $\gamma -$ ray observatories will play critical role in searches
for neutralinos with mass larger than $\sim 100$ GeV.

While direct detection and accelerator searches have an exciting discovery
potential, it should be emphasized that there is a large region of parameter
space for which gamma-ray instruments could provide the only detection for
cases where the nuclear recoil cross-section falls below the threshold of any
planned direct detection experiment, or the mass is out of range of the LHC or
even the ILC.  {Any comprehensive scientific roadmap that puts the discovery of
dark matter as its priority must include support for a future, high-sensitivity
ground-based gamma-ray experiment in addition to accelerator and direct
searches}

But the next 5-10 years of DM research may provide us with a large amount of
experimental results coming from LHC, direct DM searches
\cite{aprile05,klapdor02,klapdor05,bisset07,akerib05,sanglard07} and indirect
observations of astrophysical $\gamma$-rays.  Current gamma-ray experiments
such as AGILE, Fermi, VERITAS, HESS and MAGIC will continue making observations
of astrophysical sources that may support very high density dark matter spikes
and may, with luck, provide a first detection of dark matter.  The wide
field-of-view Fermi instrument could provide serendipitous detections of
otherwise dark, dark matter halos, and search for the unique dark matter
annihilation signal in the isotropic cosmological background.  EAS experiments
will provide evidence about the diffuse galactic background at the highest
energies, helping to understand backgrounds for dark matter searches and even
offering the potential for discovery of some unforeseen very high mass,
nonthermal relic that form the dark matter.  All of these results will guide
the dark matter research which can be conducted by a future ground-based
observatory needed to study the dark matter halos, and would affect strongly
the design parameters of such an observatory.

To briefly summarize the interplay between the LHC, Fermi and a future
ground-based gamma-ray instrument, it is necessary to consider several
different regimes for the mass of a putative dark matter particle:

\begin{itemize}

\item {\sl Case I:} If $m_\chi\sim  100\, {\rm GeV}$ and the LHC sees the LSP,
Fermi will probably provide the most sensitive measurements of the continuum
radiation and will be needed to demonstrate that a supersymmetric  particle
constitutes the dark matter \cite{koushiappas_santafe06}.  Ground-based
measurements will be needed to constrain the line-to-continuum ratio to better
determine the supersymmetric parameters or to obtain adequate photon statistics
(limited by the $\sim$m$^2$ effective area of Fermi) to obtain the smoking gun
signature of annihilation by observing line emission. 

\item {\sl Case II:} If $100\, {\rm GeV} <  m_\chi < 300\,  {\rm GeV}$, the LHC
could still see the neutralino, but both the line {and continuum emission}
could be better detected with a a low-threshold (i.e., 20-40~GeV threshold)
ground-based experiment than with Fermi, if the source location is known.
Again these gamma-ray measurements are still required to demonstrate that a
supersymmetric particle constitutes astrophysical halos, and to further measure
supersymmetric parameters \cite{baltz_p5_06}.

\item {\sl Case III:} If $m_\chi > 300\, {\rm GeV}$ future direct-detection
experiments and ground-based gamma-ray experiments may be able to detect the
neutralinos.  Only ground-based instruments will be able to determine the halo
parameters, and will provide additional constraints on SUSY parameter space
somewhat orthogonal to the constraints provided by the determination of the
direct detection cross-sections. For a sizeable fraction of parameter space,
nuclear recoil cross-sections may be too small for direct detection but the
total annihilation cross section could still be large enough for a gamma-ray
detection.  Detection at very high energies would be particularly important for
non-SUSY dark matter candidates such as the lightest Kaluza-Klein partner,
where current constraints put the likely mass range above the TeV scale.  Since
TeV-scale neutralinos are likely to be either pure Higgsino or pure Wino
particles, particle-physics uncertainties are expected to be smaller in this
VHE energy regime.
\end{itemize}

\section{Conclusions}

A next-generation $\gamma$-ray telescope has the unique ability to make the
connection from particles detected in the laboratory to the dark matter that
dominates the density of matter in the universe, and to provide important
constraints that help to identify the nature of the dark matter particle.  The
main findings of our study about the potential impact of gamma-ray measurements
on the dark-matter problem and the requirements for a future instrument are
summarized below:

\begin{itemize}

\item Compared with all other detection techniques (direct and indirect),
$\gamma$-ray measurements of dark-matter are unique in going beyond a detection
of the local halo to providing a measurement of the actual distribution of dark
matter on the sky.  Such measurements are needed  to understand the nature of
the dominant gravitational component of our own Galaxy, and the role of dark
matter in the formation of structure in the Universe.

\item There are a number of different particle physics and astrophysical
scenarios that can lead to the production of a gamma-ray signal with large
variations in the total flux and spectral shape. The spectral form of the
gamma-ray emission will be universal, and contains distinct features that can
be connected, with high accuracy, to the underlying particle physics.

\item The annihilation cross-section for gamma-ray production from higher
energy (TeV) candidates are well constrained by measurements of the relic
abundance of dark matter, with the particle-physics uncertainty contributing
$\sim$ one order of magnitude to the range of the predicted gamma-ray fluxes. 

\item The Galactic center is predicted to be the strongest source of gamma-rays
from dark matter annihilation but contains large astrophysical backgrounds.  To
search for gamma-ray emission from dark-matter annihilation in the Galactic
center region, the requirements for the future instrument include: extremely
good angular resolution to reject background from other point sources, a
moderately large field of view ($\gsim 7^\circ$ diameter), a good energy
resolution ($\lsim 15$\%), a low energy threshold $\lsim 50$~GeV, and location
at a southern hemisphere site.

\item Observations of local-group dwarf galaxies may provide the cleanest
laboratory for dark-matter searches, since these dark-matter dominated objects
are expected to lack other astrophysical backgrounds.  For these observations,
a very large effective area and excellent point-source sensitivity down to
$\lsim$50 GeV is required.  Energy resolution better than 15-20\% is required
to determine the spectral shape.  Currently, the best strategy for detecting
dark matter from dwarf galaxies, globular clusters or local group galaxies is
to observe an ensemble of sources, taking advantage of the source-to-source
variance in the halo profile that may lead to large enhancements in the signal
from some sources, although improvements in constraints on the dark-matter
density profile from future detailed astronomical measurements (e.g., from
stellar velocity dispersion) will allow for a refinement of the list of most
promising targets.

\item Observations of halo-substructure could provide important new constraints
on CDM structure formation, providing information on the mass of the first
building blocks of structure, and on the kinetic decoupling temperature.  The
most direct experimental way to probe the presence of otherwise dark halo
substructure in the Milky Way is through $\gamma$-ray observations.
Space-based low-threshold all-sky measurements will be most effective for
identifying candidate objects, but ground-based measurements will be required
to determine the detailed spectral shape (cutoff, line-to-continuum ratio)
needed to identify the dark matter candidate.

\item {The spiky halos around galactic intermediate-mass black holes could
provide a large enhancement in the gamma-ray signal that could be effectively
detected by all-sky low-threshold instruments such as Fermi or a future
space-based instrument, then followed-up by ground-based measurements.  Over
most of the allowed parameter space, Fermi would detect the onset of the
continuum spectrum but would lack the sensitivity to measure the detailed
spectral shape above hundreds of GeV.  Ground-based measurements with good
point-source sensitivity, and good energy resolution (10-15\%) would be
necessary to follow-up these detections to measure the spectral cutoff and
other features of the annihilation spectrum needed to clearly identify a
dark-matter origin of the gamma-ray signal.} 

\item {While a space-based instrument or future IACT arrays are probably the
only means of providing the large effective area, low threshold, energy and
angular resolution for detailed measurements of gamma-rays from dark matter
annihilation, future EAS experiments like HAWC can also play a useful role.
Future EAS experiments, with their wide field of view and long exposure time,
also have the potential for serendipitous discovery of some corners of
parameter space, in particular for nonthermal relics and mass close to the
unitarity limit.   The good sensitivity of EAS experiments can provide
important measurements of diffuse, hard-spectra galactic backgrounds.}  

\item {Gamma-ray astronomy has the unique potential to provide important
constraints on the history of structure formation in the universe through
dark-matter observations of dark-matter halos on a range of mass scales
(including IMBH halos) in addition to probing the history of star formation
through measurements of the diffuse infrared background radiation.} 

\item {In general, a combination of laboratory (LHC, ILC) detection and
astrophysical observations or direct detection experiments will be required to
pin down all of the supersymmetric parameters and to make the complete case
that a new particle observed in the laboratory really constitutes the dark
matter.}

\item {Gamma-ray astronomy is unique in that the detection cross-section is
closely related to the total annihilation cross section that determines the
relic abundance.}

In closing, we reiterate that a comprehensive plan for uncovering the nature of
dark matter must include gamma-ray measurements.  With an order of magnitude
improvement in sensitivity and reduction in energy threshold, a future IACT
array should have adequate sensitivity to probe much of the most generic
parameter space for a number of sources including Galactic substructure, Dwarf
galaxies and other extragalactic objects.

\end{itemize}


\begin{thebibliography}{}
\setlength{\itemsep}{-0.0em}
\setlength{\parsep}{4pt}
\setlength{\baselineskip}{13pt}


\bibitem{Spergel:2006hy}
  Spergel, D.N., {\it et al.}  [WMAP Collaboration],
  Astrophys.\ J.\ Suppl.\  {\bf 170}, 377 (2007).

\bibitem{Percival:2006gt}
  Percival, W.J.,  { et al.},
  Astrophys.\ J.\  {\bf 657}, 645 (2007)

\bibitem{baltz_p5_06} Baltz, E.A., P5 presentation (2006)

\bibitem{Zwicky1933} Zwicky, F., Helvetica Physica Acta\ {\bf 6}, 110--127 (1933).

\bibitem{CowsikMcClelland(1973)} Cowsik, R., McClelland, J., \apj, {\bf 180}, 7 (1973).

\bibitem{clowe06} Clowe, D., et al., \apjl, {\bf 648}, (2006) L109. 

\bibitem{baltz06} Baltz, E.A., et al.,
Phys. Rev. D., {\bf 74}, 103521 (2006)

\bibitem{aprile05} Aprile, E., et al., New Astron. Rev., {\bf 49}, 289 (2005)

\bibitem{klapdor05} Klapdor-Kleingrothaus, H.V., Krivosheina, I.V., Nucl.
Phys. B. (Proc. Suppl.), {\bf 145}, 237 (2005).

\bibitem{klapdor02} Klapdor-Kleingrothaus, H.V., et al., Nucl. Inst. and
Methods Phys. Res. A, {\bf 481}, 149 (2002).

\bibitem{bisset07} Bisset, R., et al., Nucl. Phys. B. (Proc. Suppl.), {\bf 173},
164 (2007).

\bibitem{akerib05} Akerib, D.S., et al., Nucl. Inst. and Methods Phys. Res. A, {\bf 559}, 411 (2005).

\bibitem{morales99} Morales, A., TAUP99, Paris (1999).

\bibitem{sanglard07} Sanglard, V., et al., Nucl. Phys. B. (Proc. Suppl.),
{\bf 173}, 99.

\bibitem{ibarratran08} Ibarra, A., Tran, D., Phys. Rev. Lett., {\bf 100},
061301 (2008)

\bibitem{bergstrom06} Bergstr\"om, L., et al., arXiv:astro-ph/0609510v1 (2006).

\bibitem{amelino98} Amelino-Camelia, G., et~al., Nature, {\bf 393}, 319 (1998)

\bibitem{biller99} Biller, S.D., et~al., Phys.~Rev.~Lett., {\bf 83}, 2108 (1999)

\bibitem{birkedal06} ABirkedal, A., et al., 
Phys.~Rev.~D., {\bf 74}, 035002 (2006).

\bibitem{kraussnasri} Krauss, L.M., Nasri, S., Trodden, M., Phys. Rev. D, {\bf 67}, 085002 (2003).

\bibitem{zee80} Zee, A., Phys. Lett., {\bf B93}, 339 (1980);
Phys. Lett., {\bf B161}, 141 (1985).

\bibitem{baltzbergstrom} Baltz, E.A., and Bergstr\"om, L., Phys. Rev. D
{\bf 67}, 043516 (2003).

\bibitem{bringmann07} Bringmann, T., Bergstr\"om, L., Edsj\"o, J.,
arXiv:hep-ph/0710.3169v1 (2007).

\bibitem{ferrer06} Ferrer, F., Krauss, L.M., Profumo, S., Phys. Rev. D.,
{\bf 74}, 115007 (2006)

\bibitem{nfw97} Navarro, J.~F., Frenk, 
C.~S., \& White, S.~D.~M. \apj, {\bf 490}, 493 (1997).

\bibitem{moore99} Moore, B., Quinn, T., 
Governato, F., Stadel, J., \& Lake, G.\, \mnras, {\bf 310}, 1147  (1999).

\bibitem{Netal} Navarro, J.F., et~al., MNRAS, {\bf 349},
1039 (2004).

\bibitem{DZMSC} Diemand, J., et~al., MNRAS, {\bf 364}, 665
(2005).

\bibitem{SKBK06} Strigari, L.E., et al., 
astro-ph/0611925 (2006).

\bibitem{bub98} Bergstr{\"o}m, L., Ullio, P., Buckley, J.H., Astroparticle Physics, {\bf 9}, 137 (1998).

\bibitem{kosack04} Kosack, K., { et al.}, \apjl, {\bf 608}, L97 (2004).

\bibitem{GCHESS2004} Aharonian, F., { et~al.}, \aap {\bf 425}, L13--L17 (2004).

\bibitem{Horns2005} Horns, D., Physics Letters B, {\bf 607}, 225--232 (2005).

\bibitem{Profumo:2005xd} Profumo, S., Phys. Rev. D, {\bf 72}, 103521 (2005).

\bibitem{Stoehr:2003hf}
  Stoehr, F., et al.,
  MNRAS {\bf 345}, 1313 (2003)

\bibitem{serpico08}
P.D.~Serpico and G.~Zaharijas, 2008arXiv0802.3245S (2008)

\bibitem{Diemand:2006ik}
  Diemand, J., Kuhlen, M., Madau, P., 
  Astrophys.\ J.\  {\bf 657}, 262 (2007)

\bibitem{fornengo04} Fornengo, N., Pieri, L., Scopel, S.,  Phys. Rev. D.,
{\bf 70}, 103529 (2004).


\bibitem{Vassiliev2003} Vassiliev, V.V., 
28th Int. Cosmic Ray Conf.(Tsukuba), 2679-2682 (2003).

\bibitem{LeBohec2003} LeBohec, S. L.,  \& VERITAS Collaboration,
28th Int. Cosmic Ray Conf. (Tsukuba), 2521-2524 (2003).

\bibitem{woodetal07} Wood, M., et~al., submitted (2007).

\bibitem{moulin07} Aharonian, F., et~al., Astropart.~Phys., {\bf 29}, 55 (2008).

\bibitem{Betal02} Baltz, E., et~al., Phys. Rev. D, {\bf 61}, 023514 (2000).

\bibitem{Tyler02} Tyler, C., Phys. Rev. D, {\bf 66}, 023509 (2002).

\bibitem{EFS04} Evans, N.W., Ferrer, F., Sarkar, S., Phys. Rev. D, {\bf 69},
123501 (2004).

\bibitem{BH06} Bergstr\"om, L., Hooper, D., Phys. Rev. D, {\bf 73}, 063510 (2006).

\bibitem{PB04} Pieri, L., Branchini, E., Phys. Rev. D, {\bf 69}, 043512 (2006).

\bibitem{Strigari:2007at}
  Strigari, L.E., et al.
  arXiv:0709.1510 [astro-ph].

\bibitem{belokurov} Belokurov, V.,  et al., Astrophys.~J., {\bf 654}, 897 (2007).

\bibitem{martin07} Martin, N.F., et al., MNRAS, {\bf 380}, 281 (2007).

\bibitem{metz} Metz, M., Kroupa, P.,  MNRAS, {\bf 376}, 387 (2007).

\bibitem{munoz07} Munoz, R.R., Majewski, S.R., Johnston, K.V., arXiv:0712.4312v1 [astro-ph] (2007).

\bibitem{Klypin:1999uc} Klypin, A., et al., 
Astrophys. J. {\bf 522}, 82 (1999).

\bibitem{Moore:1999wf} Moore, B., et~al., Astrophys. J., {\bf 524}, L19 (1999).

\bibitem{1992ApJ...398...43C} Carlson, E.D., Machacek, M.E., Hall, L.J., 
Astrophys. J., {\bf 398}, 43 (1992).

\bibitem{Spergel:1999mh} Spergel, D.N., Steinhardt, P.J., Phys. Rev. Lett.,
{\bf 84}, 3760 (2000).

\bibitem{Kaplinghat:2000vt} Kaplinghat, M., Konx, L., Turner, M.S., Phys. Rev.
Lett., {\bf 85}, 3335, (2000).

\bibitem{Kamionkowski:1999vp} Kamionkowski, M., Liddle, A.R.,  Phys. Rev. Lett.,
{\bf 84}, 4525 (2000).

\bibitem{Zentner:2002xt} Zentner, A.R., Bullock, J.S., Phys. Rev. {\bf D66},
043003, (2002).

\bibitem{1986ApJ...303...39D} Dekel, A., Silk, J., Astrophys. J., {\bf 303}, 39
(1986).

\bibitem{1999ApJ...523...54B} Barkana, R., Loeb, A., Astrophys. J., {\bf 523},
 54, (1999).

\bibitem{2001ApJ...548...33B} Bullock, J.S., Kravtsov, A.V., Weinberg, D.H., 
Astrophys. J., {\bf 548}, 33, (2001).

\bibitem{Koushiappas:2003bn} Koushiappas, S.M., Zentner, A.R., Walker, T.P., 
Phys. Rev., {\bf D69}, 043501, (2004).

\bibitem{Setal99} Schmid, C., et~al., Phys. Rev. D, {\bf 59}, 043517 (1999).

\bibitem{Hetal01} Hofmann, S., et~al., Phys. Rev. D, {\bf 64}, 083507 (2001).

\bibitem{Chenetal01} Chen, X., et~al., Phys. Rev. D, {\bf 64}, 021302 (2001).

\bibitem{Berezinskyetal03} Berezinsky, V., et~al., Phys. Rev. D, {\bf 68},
103003 (2003).

\bibitem{Green04} Green, A.M., Hofmann, S., Schwarz, D.J., MNRAS, {\bf 353}, L23 (2004).

\bibitem{Green05} Green, A.M., et~al., J. Cosmol. Astropart. Phys., {\bf 08},
003, (2005).

\bibitem{LZ05} Loeb, A., Zaldarriaga, M., Phys. Rev. D, {\bf 71}, 103520 (2005).

\bibitem{Bertch} Bertschinger, E., Phys. Rev. D, {\bf 74}, 063509 (2006).

\bibitem{Mooreetal06} Moore, B., et~al., astro-ph/0502213

\bibitem{K06} Koushiappas, S.M., Phys. Rev. Lett., 97, 191301 (2006).

\bibitem{peebles:1972} Peebles, P. J. E.,  Astrophys.\ J.\, {\bf 178}, 371 (1972).

\bibitem{young:1980}
  Young, P., Astrophys.\ J.\ {\bf 242}, 1232 (1980).

\bibitem{Ipser:1987ru}
  Ipser, J.R., Sikivie, P.,
  Phys.\ Rev.\ D {\bf 35}, 3695 (1997).

\bibitem{Quinlan:1995}
  Quinlan, G.D., Hernquist, L., Sigurdsson, S.,
  Astrophys.\ J.\  {\bf 440}, 554 (1995).

\bibitem{Gondolo:1999ef}
  Gondolo, P., Silk, J., 
  Phys.\ Rev.\ Lett.\  {\bf 83} 1719 (1999)

\bibitem{Merritt:2003qc}
  Merritt, D.,
  Proceedings of Carnegie Observatories Centennial Symposium
  [arXiv:astro-ph/0301257].

\bibitem{Bertone:2005hw}
  Bertone, G., Merritt, D.,
  Phys.\ Rev.\ D {\bf 72}, 103502 (2005)

\bibitem{Bertone:2005xv}
  Bertone, G., Merritt, D.,
  Mod.\ Phys.\ Lett.\ A {\bf 20}, 1021 (2005)

\bibitem{Ullio:2001fb}
  Ullio, P., Zhao, H.,  M.~Kamionkowski, M.,
  Phys.\ Rev.\ D {\bf 64}, 043504 (2001)

\bibitem{Merritt:2002vj}
  Merritt, D., Milosavljevic, M., Verde, L., Jimenez, R.,
  arXiv:astro-ph/0201376.

\bibitem{Merritt:2003eu}
  Merritt, D., 
  arXiv:astro-ph/0301365.

\bibitem{Madau:2001}
  Madau, P., Rees,M.J., Astrophys.\ J.\ Lett.\, {\bf 551}, L27 (2001).

\bibitem{Bertone:2005xz}
  Bertone, G., Zentner, A.R., Silk, J.,
  Phys.\ Rev.\ D {\bf 72}, 103517  (2005).

\bibitem{Zhao:2005zr}
  Zhao, H.S., Silk, J.,
  arXiv:astro-ph/0501625.

\bibitem{islamc:2004}
  Islam, R., Taylor, J., Silk, J.,
  MNRAS {\bf 354}, 443 (2003).

\bibitem{islamb:2004}
  Islam, R., Taylor, J., Silk, J.,
  MNRAS {\bf 354}, 427 (2004).

\bibitem{Koushiappas:2003zn}
  Koushiappas, S.M., Bullock, J.S., Dekel, A.,
  MNRAS\  {\bf 354}, 292 (2004)

\bibitem{JE2004} Edsj{\"o}, J., Schelke, M., Ullio, P.,
Journal of Cosmology and Astro-Particle Physics, {\bf 009}, 004-1--004-25
  (2004).

\bibitem{BG2002} Baltz, E.A., Gondolo, P.,
{\prd}, {\bf 67}, 063503-1--063503-8 (2003).

\bibitem{Bottino2003} Bottino, A., Donato, F., Fornengo, N., Scopel, S.,
{\prd}, {\bf 68}, 043506-1 (2003)

\bibitem{Bottino2004} Bottino, A., Donato, F., Fornengo, N., Scopel, S.,
{\prd}, {\bf 69}, 037302-1 (2004)

\bibitem{GK1990} Griest, K., Kamionkowski, M.,
{\prl}, {\bf 64}, 615 (1990).

\bibitem{profumo05}  Profumo, S., \prd, {\bf 72}, 
103521 (2005).

\bibitem{tsuchiya04} Tsuchiya, K., Enomoto, R., Ksenofontov, L. T., et al., ApJ 606, L115 (2004)..

\bibitem{koushiappas_santafe06} Koushiappas, S., talk at
"Ground-based Gamma-ray Astronomy: Towards the Future, May 11-12, Santa Fe (2006).


\end{thebibliography}
\end{document}